\documentclass[lettersize,journal]{IEEEtran}
\usepackage{amsmath,amsfonts}
\usepackage{algorithmic}
\usepackage{algorithm}
\usepackage{array}
\usepackage{textcomp}
\usepackage{stfloats}
\usepackage{url}
\usepackage{verbatim}
\usepackage{graphicx}
\usepackage{cite}
\usepackage{amsmath,amssymb,amsfonts}
\usepackage{algorithmic}
\usepackage{graphicx}
\usepackage{textcomp}
\usepackage{algorithm,algorithmic}
\usepackage{mathtools}
\DeclarePairedDelimiter\ceil{\lceil}{\rceil}

\usepackage{xcolor}
\usepackage{graphicx}
\usepackage{multirow}
\usepackage{float}
\usepackage{amsmath}
\usepackage{algorithm,algorithmic}
\usepackage{cite}
\usepackage{amsmath,amssymb,amsfonts}
\usepackage{algorithmic}
\usepackage{graphicx}
\usepackage{textcomp}
\usepackage{xcolor}
\usepackage{soulutf8}
\usepackage{pifont}
\newcommand{\cmark}{\ding{51}}%
\newcommand{\xmark}{\ding{55}}%

\begin{document}
%Believe there are NO Limits but the Energy:
\title{An Efficient NVM based Architecture for Intermittent Computing under Energy Constraints}

\author{SatyaJaswanth Badri, Mukesh Saini, and Neeraj Goel
        % <-this % stops a space
\thanks{The authors are with the Computer Science and Engineering Department, IIT Ropar, Punjab-140001, India (e-mail: 2018csz0002@iitrpr.ac.in; mukesh@@iitrpr.ac.in; neeraj@iitrpr.ac.in).}}% <-this % stops a space
%\thanks{Manuscript received April 19, 2021; revised August 16, 2021.}}



\maketitle

\begin{abstract}
Battery-less technology evolved to replace battery technology. Non-volatile memory (NVM) based processors were explored to store the program state during a power failure. The energy stored in a capacitor is used for a backup during a power failure. Since the size of a capacitor is fixed and limited, the available energy in a capacitor is also limited and fixed. Thus, the capacitor energy is insufficient to store the entire program state during frequent power failures. This paper proposes an architecture that assures safe backup of volatile contents during a power failure under energy constraints. Using a proposed dirty block table (DBT) and writeback queue (WBQ), this work limits the number of dirty blocks in the L1 cache at any given time. We further conducted a set of experiments by varying the parameter sizes to help the user make appropriate design decisions concerning their energy requirements. The proposed architecture decreases energy consumption by 17.56\%, the number of writes to NVM by 18.97\% at LLC, and 10.66\% at a main-memory level compared to baseline architecture.

\end{abstract}


\section{Introduction}

The Internet of Things (IoT) allows humans to interact and connect with almost every object. These wearable and implantable devices consist of many sensors. In the future, IoT may consist of billions of sensors and systems by the end of 2050 \cite{big}. Most of these devices will be powered by batteries. Maintaining and replacing a larger number of battery-operated devices is a costly and enormous task. Further, batteries are hazardous to the environment, and their lifetime is also a critical issue \cite{lifetime}.

The alternative solution is to harvest energy from the environment and use it in the IoT system, thus, completely avoiding the use of batteries. Energy harvesting is unpredictable, and power failures are often \cite{energy12}. Thus, devices based on harvested energy can also be referred to as intermittently powered devices or battery-less devices \cite{int3, intermittent}. In conventional processors, registers, cache, and main memories are volatile \cite{forget}; therefore, after each power failure, all data is erased from the memory, and the system has to restart from the beginning.

Non-volatile processors (NVP) have been proposed \cite{r8, r41, forget} as a solution in the past. An NVP stores the processor state in non-volatile memory (NVM) during a power failure. Thus, NVP resumes the application's computational tasks once the power supply is restored, thus achieving faster recovery and backup speeds when compared to traditional processors.

At power failure, NVP needs to store the content of volatile memory (registers/SRAM caches) in an NVM such that the application can restart from the same point. The size of the registers and the contents of the SRAM caches determines the time and energy required for backup. Although using no SRAM caches reduces backup time and energy to almost negligible, but it significantly impacts performance.

When using battery-less hardware, the device must be turned off as soon as harvested power is no longer available. To avoid sudden power failures and fluctuations, such devices accumulate energy in a capacitor that smoothens the power availability and provides energy during power failures \cite{capac, int2,int3}. Thus, during a power failure, the energy stored in a capacitor is used to backup the processor state. The entire process state in volatile memory must be backed up to ensure correctness. Further, cache lines store a copy of data elements present in memory; therefore, cache lines that are not modified need not be backed up. However, in the worst case, all the cache lines could be dirty.

Since the energy storage capacity of the capacitor is limited and fixed, only fixed SRAM contents can be copied to NVM during a power failure. A sub-optimal solution is to constrain the cache size based on the energy available in a capacitor or design the capacitor to store the entire cache.

This paper proposes an NVM-based architecture that can save the process state by optimally utilizing a fixed amount of energy and using effective cache management policies. We propose to use a cache larger than that can be backed up by the capacitor. To limit the number of cache lines to be backed up, we fix the maximum number of dirty blocks in the cache. We propose a cache architecture that maintains a maximum number of dirty blocks and cache management policies that keep track of dirty blocks and suggest which block to replace.

The proposed architecture is compared with write-back and write-through cache architectures in terms of performance and energy consumption. During stable power supply, the proposed architecture reduces energy consumption by 17.56\%, writes to STT-RAM by 18.97\%, and PCM by 10.66\% compared to baseline architecture. During power failure, the proposed architecture consumes 20.94\% less energy than the baseline.

\textbf{Paper organization:} Section \ref{p21} discusses the background and related works. Section \ref{system} explains the motivation behind the base architecture selection and the need to propose an NVM-based architecture that uses only a fixed amount of energy for backup. Section \ref{p4} describes the proposed architecture in detail. The experimental setup and results are discussed in section \ref{exp}. We conclude this work in section \ref{p6}.

\section{Background and Related Works} \label{p21}

SRAM/DRAM is used to design registers, caches, and main memory for conventional processors. Recent advancements in NVM technologies include Spin-Transfer Torque RAM (STT-RAM) \cite{stt1}, Phase Change Memory (PCM) \cite{pcm}, and Ferroelectric RAM (FRAM) \cite{msp}. These NVM technologies motivated researchers for their appealing characteristics, such as non-volatility, low cost, and high density. NVM is used to design flip-flops \cite{rev2,reg2}. S. Thirumala et al. \cite{rev2} proposed reconfigurable-ferroelectric transistors to design energy-efficient intermittent devices. NVM is used to design the L1 cache \cite{xie}, last level cache (LLC) \cite{llc1,llc2}, and main memory \cite{msp,pcm} in the literature. Many studies have used NVMs to build even hybrid memories \cite{stt1,choi,xie}, saving significant energy when configured and used correctly.

Writes to NVMs consume more latency and energy compared to volatile memory. We must optimize NVM utilization by reducing the usage of NVM or reducing the number of writes to NVM. Many researchers are working to reduce the number of NVM writes at the cache or main memory. Choi et al. \cite{choi} proposed a way allocation scheme to reduce write counts to NVM in their hybrid LLC. Lee et al. \cite{r2} introduced PCM buffers to overcome the overheads, i.e., write latency and energy. Qureshi et al. \cite{r3} proposed a write cancellation and write pausing technique to give more priority to read requests than write requests. 

% Hybrid main memory architectures \cite{r5,r7} have been introduced to efficiently use DRAM and PCM for reducing write latency and energy. 

To develop an intermittent aware design, we should also change the execution model of a conventional processor by incorporating additional backup/restore procedures \cite{intermittent}. We require an efficient backup/restore procedure that backup and restores volatile contents during power failures. The size of volatile contents determines the amount of energy required to backup/restore during a power failure \cite{backup}. If we only have a small amount of available energy in our capacitor, and this energy is insufficient to backup the entire volatile contents, we may get inconsistent results. As a result, we must reduce backup/restore overheads during frequent power failures.

Many researchers are working to reduce NVP backup and restore overheads. Recent checkpointing techniques \cite{check2,check3,xie,rev2} are also proposed to reduce the size of volatile contents that must be backup/restored during frequent power failures. In-situ checkpointing has gained popularity in recent times \cite{rev2,rev21}, which uses unified NVM architectures. Lee et al. \cite{r8} proposed an adaptive NVP that prioritizes data retention to reduce the frequency of backup/restore operations. The number of power failures can be reduced by voltage and frequency scaling \cite{r9}, \cite{r10}. Rather than reducing the number of backups and restore operations, researchers \cite{r11} focused on reducing the size of backup contents. Architectures \cite{r12}, \cite{r13} based on the comparison and compression strategies are proposed to reduce the number of bits/contents to be stored in the non-volatile flip flop (NVFF)-based NVP, which reduces the dynamic energy consumption.

\section{Motivation} \label{system}

This section discusses the observations that motivated us to propose an energy-efficient NVM-based architecture.

\subsection{\textbf{Motivation for base architecture}} \label{mot11}

\textbf{NVMs at Cache Level:} Typically, we want LLCs to be small and faster in comparison to main memory. Based on these two parameters, we must choose between STT-RAM and PCM to determine which NVM technology is appropriate for LLC. For small-size LLCs, STT-RAM outperforms PCM because STT-RAM consumes less latency and low energy and has high write endurance than the PCM. In terms of write energy and write latency, PCM consumes 10 times more than STT-RAM. The endurance of STT-RAM is $4 \times 10^{12}$ write cycles, whereas PCM has $10^9$ write cycles \cite{cache2n}. When used at the cache level, STT-RAM has a lifetime of more than tens of years \cite{cache2n}. As a result, we use STT-RAM as a replacement for SRAM at the cache level. Thus, we use STT-RAM at LLC throughout the paper.

\textbf{NVMs at Main Memory Level:} Usually, we want our main memories to be large and inexpensive. Based on these two parameters, we must select one of several NVM technologies, including STT-RAM, PCM, flash, and FRAM. NAND flash has the fewest write/erase cycles. A block must be erased every time in a NAND flash before writing to it, which consumes extra energy and delay \cite{n3,n11,n26}. FRAM cannot yet replace either DRAM or NVM technologies in terms of density because of its scalability \cite{n108}. Though PCM is slower than STT-RAM, but PCM has better density characteristics. PCM has 4x times more density than STT-RAM \cite{cache1n}. STT-RAM is more expensive than PCM. PCM is 2x-4x slower than DRAM but provides 4x more density than DRAM \cite{pcm}. The majority of previous works have used PCM as an emerging candidate at the main memory level \cite{r1}, \cite{r2}, \cite{r3}. PCM is a viable alternative to DRAM in main memory design. Thus, we use PCM as the main memory throughout the paper.

\begin{figure*}[htb]
  \includegraphics[width= 1\linewidth]{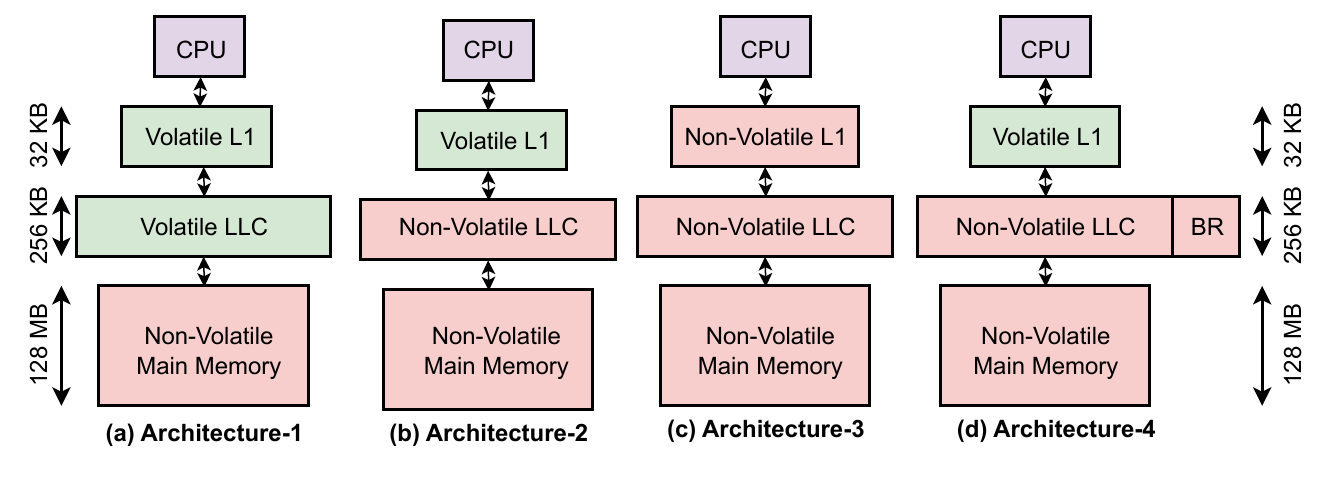}
 \caption{Architecture Designs to integrate NVM (a) Introduce NVM at Main-memory, and (b) Introduce NVM at the last-level cache and main-memory levels, (c) Introduce NVM at both the cache levels and main-memory level, and (d) Introduce NVM based Backup-Region (BR) at the last-level cache.}
 \label{fig2d2}
 \end{figure*}

\textbf{Evaluating the base architecture :} We chose NVM for LLC and main memory based on previous discussions. We use STT-RAM at LLC and PCM at main memory throughout this work. In architecture-1, SRAM is used at L1 and LLC, and PCM is used at the main memory, as shown in figure \ref{fig2d2} (a), i.e., traditional architecture. Architecture-2  is shown in figure \ref{fig2d2} (b), SRAM is used at L1, STT-RAM at LLC, and PCM is used at the main memory, i.e., our base architecture. These comparisons help to evaluate how bad or good our base architecture is under stable and unstable power scenarios.

\textbf{Under stable power supply :} Both architecture-1 and 2 give an equal number of writes to PCM during a regular operation. In architecture-2, STT-RAM takes more cycles to execute than in architecture-1 because the CPU has to stall to complete each STT-RAM write. So, we implement the LLC cache such that the LLC gets less number of writes by passing the writes to PCM to hide the LLC latency. We observed that architecture-2 takes 5.88\% more execution time than architecture-2 during stable power. Thus, using STT-RAM at the cache should be a minimal impact on overall system performance.

\textbf{Under unstable power supply :} During frequent power failures, writeback of volatile contents is essential. In architecture-1, PCM gets more writes due to frequent power failures, which consumes more energy. In architecture-2, PCM gets fewer writes because STT-RAM can save cache blocks during a power failure, which consumes less energy than in architecture-1. We compared architecture-1 and architecture-2 in terms of energy and performance during frequent power failures. We used the same set of benchmarks used in section \ref{exp} and the same experimental setup shown in table \ref{tab1}. We observed that architecture-1 takes 8.13\% more execution time than architecture-2 during power failures. On average, architecture-2 saves energy of 0.07\% per every power failure. If the number of power failures is 200, then we save 9.04\% of the overall system energy. 

Thus, using NVM at both LLC and main memory saves energy during frequent power failures. Therefore, we choose architecture-2 as our base architecture throughout this work.

\subsection{\textbf{Fixed backup energy}} \label{MK}

We need to back up the SRAM dirty blocks during a power failure to NVM. In architecture-2, we require a capacitor that helps to backup the entire L1 dirty blocks to either PCM or STT-RAM. Equation \ref{mot2} formulates the backup energy required for architecture-2. In the worst case, we need to backup the entire L1 contents to either STT-RAM or PCM.

\begin{equation} \label{mot2}
E_{backup/A2} =  N_{B/L1}  \times (e_{w\_sttram})
\end{equation}

Where $E_{backup/A2}$ is the backup energy required for architecture-2 during a power failure. $N_{B/L1}$ is the number of blocks at L1, and $e_{w\_sttram}$ is the energy per write for the STT-RAM cache block.

Usually, a capacitor has fixed energy $(E_{capacitor})$ that can only backup a fixed number of cache blocks $(K)$ during a power failure. In the worst case, we need to backup the entire L1 contents to either STT-RAM or PCM for architecture-2. We need a larger capacitor to backup the entire L1 contents, as shown in equation \ref{mot2}, which is infeasible in practice. A large capacitor requires more time to charge. Thus, maintaining the larger size capacitor will not help us during frequent power failures, resulting in faulty computations. This observation motivated us to propose an architecture that uses fixed energy to backup the L1 dirty cache contents during a power failure.

We defined $E_{capacitor}$ in equation \ref{eq5}. Where C is the capacitance, and V is the operating voltage.

\begin{equation} \label{eq5}
 E_{capacitor} = \frac{1}{2} C V^2
\end{equation}

So, given $E_{capacitor}$ as constant, our objective is to maximize $K$ cache blocks. We define $K$ using equation \ref{eq4}. 

\begin{equation} \label{eq4}
  K = \frac{E_{capacitor} - E_{reg\_file}}{e_{w\_sttram}} 
\end{equation}

Where $K$ is the maximum number of blocks that can backup to NVM during a power failure. Where $E_{reg\_file}$ is the energy required to backup the register file to STT-RAM.

Instead of saving entire SRAM contents during a power failure, we backup only $K$ blocks from the L1 cache to NVM. Where $K<<\left(N_{dirty/L1}\right)$ w.r.t architecture-2 during a backup procedure.

Thus, we require a capacitor of size $C$ that can provide energy of $\ge E_{capacitor}$; this ensures that $K$ blocks are completely and safely backed up to NVM. We assumed that the capacitor energy was fixed and could not be replaced again. As a result, $E_{capacitor}$ is sufficient to backup $K$ blocks from the L1 cache and the register file contents to NVM, which we explained in section \ref{PF}.

\section{Proposed Architecture} \label{p4}
This section explains the proposed architecture. The main objective here is to use the given $E_{capacitor}$ efficiently and complete the backup within the given $E_{capacitor}$ during a power failure. We need to restrict the number of dirty blocks to $K$ to achieve the above objective. We need to address the following issues that help to restrict the number of dirty blocks to $K$:

\begin{enumerate}
    \item  The number of dirty blocks at any point in time needs to be counted and tracked.
    \item The write time to LLC is longer than the L1 cache. Every $(K+1)th$ dirty block would require additional time to write back to LLC. The processor would stall during this time, which degrades the system's performance. How can we avoid this unnecessary stalling?
 \item We need to decide which block should be replaced when dirty blocks are more than $K$.
 \item Where should all dirty blocks be stored during a power failure? 
\end{enumerate}
  
Our proposed architecture addresses these four issues. In the proposed architecture, we divided $K$ as $M+N$ blocks. To address the first issue, we proposed a dirty block table (DBT) that tracks $M$ dirty blocks. We discussed DBT in section \ref{DBT}. We introduced a write-back queue (WBQ) at the L1 cache, which resolves the second issue. We discussed WBQ in the section \ref{DBT}. To address the third issue, we explore two different replacement policies. These replacement policies are discussed in section \ref{RP}. We introduced an STT-RAM-based backup region at LLC, which provides additional storage space for volatile data during power failures.

\subsection{Proposed Architecture} \label{DBT}

\begin{figure}[htb]
  \includegraphics[width= 0.95\linewidth]{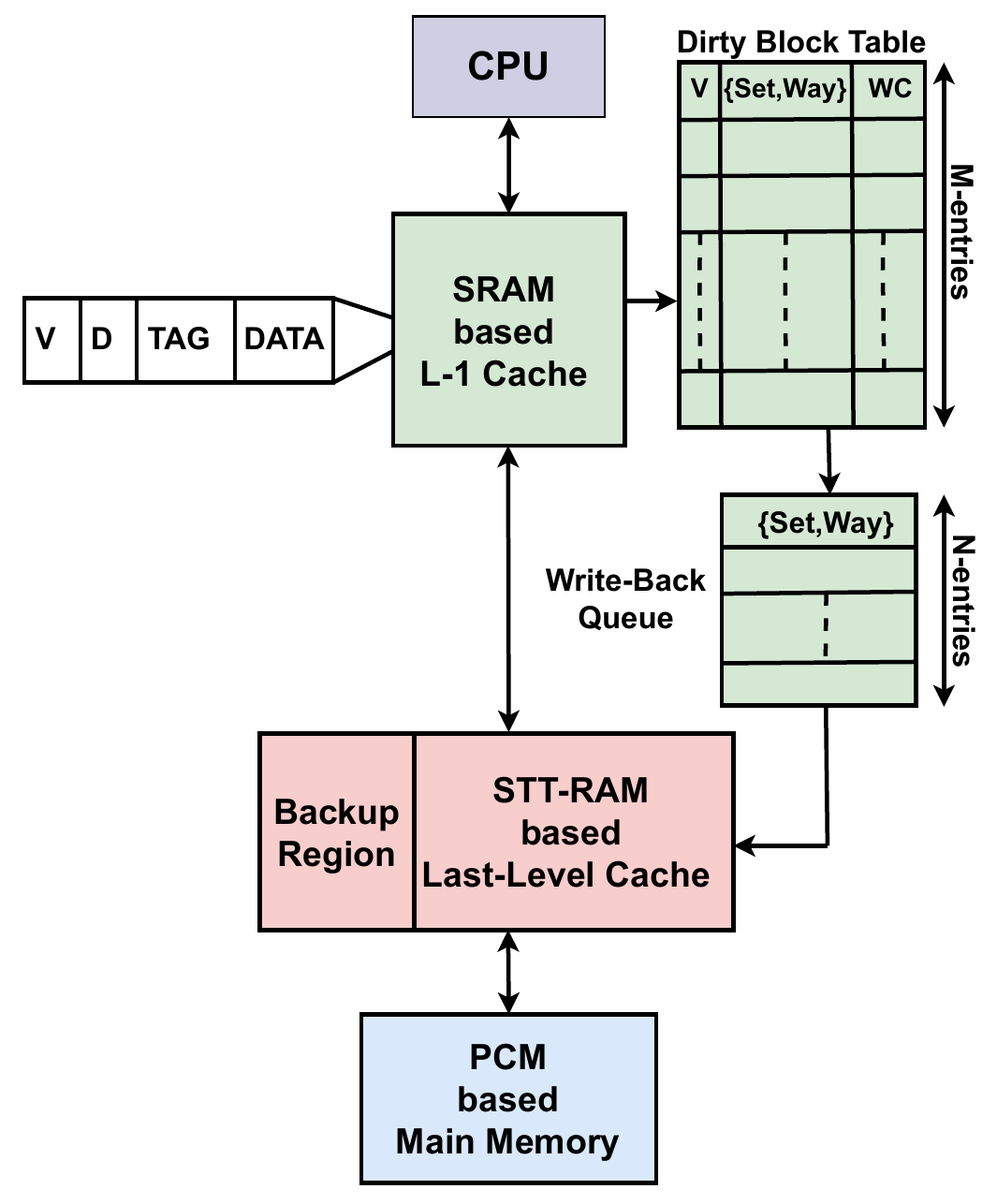}
\caption{Overview of Proposed Architecture}
\label{md}
\end{figure}

The proposed architecture is shown in figure \ref{md}. Each cache block contains a valid bit (V), a dirty bit (D), a tag, and data. 

Algorithm-\ref{alg1} refers to whenever there is a write hit at the L1 cache. Line 1 checks; if a write hit occurs and the dirty bit is 0, we set the dirty bit to 1 and create an entry in DBT. Line 3 checks whether the number of valid DBT blocks equals M. If the number of valid blocks in DBT equals M, we use the DBT replacement policy to make space for the new entry. Line 6 checks if there are more than N entries in WBQ; the processor stalls to complete a writeback to LLC. Line 12 determines whether the number of valid blocks in DBT is less than M and inserts an entry into DBT. Line 15 determines whether there is a write request and if the dirty bit is already set to 1. If Line 15 becomes true, we update the WC field in DBT.

If we find any dirty block, we make an entry to DBT. DBT stores dirty block information in four fields: valid bit (V), set, way, and write counter (WC). DBT is implemented as an M-entry fully associative buffer. DBT doesn't impact the clock period because it doesn't come under the critical path. When a victim entry is chosen from DBT for replacement, the tag information of that entry is moved into WBQ. 

WBQ has N entries that store the \{set, way\} field information. To hide the latency of STT-RAM writes, we use a writeback queue in a standard mechanism. WBQ works as a queue and writes the data to LLC to maintain N entries. When data is written from WBQ, the dirty bit in the cache is cleared. We update the modified value in WBQ whenever there is a write hit to the WBQ entry. 

The primary importance of WBQ is seen in a scenario where we need to write back the data to LLC, which takes an additional number of cycles for every $(K+1)th$ block. During this time, the processor would halt for the `X' number of cycles to write one of the $(K+1)th$ blocks to LLC, where X is the number of cycles required to complete one STT-RAM write. This additional stalling degrades system performance and consumes a significant amount of energy.

Rather than saving the entire contents of the SRAM, we backup $M+N$ dirty blocks to the STT-RAM. This reduces writes to STT-RAM and PCM during a power failure. In case of a miss at the L1 cache, the architecture is the same as conventional architecture. 

\begin{algorithm}[htp]
{\fontsize{9pt}{9pt}
\selectfont 
\caption{L1 cache hit access}
On access to block b in set s
\label{alg1}
\begin{algorithmic}[1]
\IF{ (b.write) $\And$ !(b.D)}
\STATE b.D =1
\IF{DBT.size() == M}
\STATE invokes replacement policy.
\STATE replacement policy returns a victim DBT entry.
\IF{WBQ.isFull()}
\STATE STALL
\ELSE
\STATE Make an entry in WBQ.
\ENDIF
\ELSE
\STATE Make an entry in DBT.
\STATE Update WC field in DBT.
\ENDIF
\ELSIF{b.write}
\STATE Update WC.
\ELSE
\STATE this is a read-hit case; provide the data.
\ENDIF
  
\end{algorithmic} 
 }
\end{algorithm}

 \subsection{Replacement Policy in DBT} \label{RP}
When the DBT size exceeds $>$M, we require a replacement policy in the DBT to replace any of the M entries. The traditional LRU replacement policy does not work for our architecture because we want to replace a block based on the number of writes or write behavior. For the proposed architecture, we explore two replacement policies. First, the least frequently written (LFW) policy replaces an entry that has received the least number of writes compared to all other entries in the DBT. Second, the least recently written (LRW) policy maintains the recency information for each block. Instead of replacing the block based on write counts, the LRW policy recommends replacing the most recently written block to preserve write access recency information.

We introduce a write counter (WC) field in the DBT to identify the LFW block; the size of the WC depends on `M'. If a write request is made to any DBT entry, we increment WC by one. For instance, the WC has a size of 5 bits. If the WC of the requested entry equals the maximum value ($2^{5}-1$), we do the logical right shift, i.e., we decrement $(2^{5}/2)$ from all DBT entries. For example, WC size is 5 bits, and DBT has four entries with the WC values as follows \{19, 17, 31, 3\}. Suppose a new write request is received for the third entry. The WC value for the third entry exceeds 31; we subtract 16 from all DBT entries, which becomes \{3, 1, 15, 0\}. We replace the entry with the lowest WC value during a replacement request.

We used the LRW field in the DBT to implement the LRW replacement policy. The LRW field has a size of $\ceil*{logM}$ bits. For example, if M is 16, we require a 4-bit LRW field. The implementation of LRW is identical to that of the 4-bit priority queue. When we use the LRW policy, we replace the WC field in DBT with the LRW field.

\subsection{During Intermittent power supply} \label{PF}
Apart from the STT-RAM-based LLC, we introduce a backup region (BR). STT-RAM is used to implement the backup region. The backup region can always have a maximum size of $K$ blocks + reg file. For reading/updating the backup region, we used the same access latency and energy values as the STT-RAM cache. During a power failure, we have registers and $K$ block (M+N) contents to backup. When the power comes back, we move the backup region contents to the L1 cache. With these contents, we begin the application's execution. 

\subsubsection{Validity of Proposed work} The following activities occur in the system during a power failure and before initiating the backup procedure.
\begin{enumerate}
    \item Except for the processor, all system peripherals receive power-off signals. 
    \item As the processor is stalled, no new instructions are carried out by the processor.
    \begin{enumerate}
    \item Since all peripherals are switched off, no peripheral can issue an interrupt during this time. 
    \end{enumerate}
    \item Because the processor is stalled, it consumes no dynamic energy. As a result, we can use the entire capacitor energy to backup the volatile contents.

\end{enumerate}

After completing the preceding three phases, we initiate the backup procedure. During backup, the volatile contents are stored in the register file and the SRAM-based cache; a specially designed controller is responsible for saving the volatile contents to STT-RAM-based BR at LLC. During backup, the controller reads the (set, way) fields in DBT and WBQ one entry after another to identify a block in the L1 cache, then writes the dirty data to the STT-RAM-based BR at LLC. The controller performs the following operations: reading from the register file, reading from the SRAM-based cache (‘K’ blocks data), and writing to STT-RAM-based BR at LLC. In the proposed architecture, the register file size and the number of dirty blocks at the L1 cache are minimal, and the backup procedure requires a constant and fixed number of cycles. As a result, the proposed architecture requires a fixed overhead for backup.

The processor has been set to shutdown mode once the backup procedure has been completed. We use the capacitor energy $(E_{capacitor})$ to perform the backup. When the power is resumed, we perform the same operations as in a conventional processor.

Therefore, for a given capacitor energy, system configuration, register files and read/write parameters for the memory system, we define the required number of $K$ blocks in section \ref{MK} using equation \ref{eq4}.

\section{Experimental Setup and Results} \label{exp}
\subsection{Experimental Setup} \label{exp1}
The proposed architecture is evaluated using the gem5 \cite{13} simulator and 14 MiBench benchmarks \cite{14}. Table \ref{tab1} shows the micro-architectural parameters used for implementation. We collected dynamic energy and latency values for a single read and write operation to SRAM and STT-RAM using Nvsim \cite{25}, as shown in table \ref{tab2}. 

\begin{table}[htb]
\centering
\caption{System Configuration}
\label{tab1}
\resizebox{\columnwidth}{!}{%
\begin{tabular}{|l|l|}
\hline
\textbf{Component}   & \textbf{Description}             \\ \hline
\textbf{CPU core}    & 1-core, 480MHZ               \\ \hline
\textbf{L1 Cache} &
  \begin{tabular}[c]{@{}l@{}}Block size is 64-byte, 4-way associative \\ Private cache \\ (16KB D-cache,and 16KB I-cache)\end{tabular} \\ \hline
  \textbf{Last-Level Cache} &
  \begin{tabular}[c]{@{}l@{}}Block size is 64-byte, 16-way associative \\ Private cache \\ (128KB D-cache, and 128KB I-cache), \\ write-back cache policy\end{tabular} \\ \hline
\textbf{Size Parameters} &
  \begin{tabular}[c]{@{}l@{}}VB - 1bit, WC - 6bits \\ K- 16; M- 12, N- 4, LRW- 4bits, and \\ C should be $\ge$ 1.92 nF\end{tabular} \\ \hline
\textbf{Main memory} & 128MB PCM \\ \hline
\textbf{Others} & \begin{tabular}[c]{@{}l@{}} Clock Period: 2ns, \\SRAM Read: 1 Cycle, \\ SRAM Write: 2 Cycles,\\ STT-RAM Read: 2 Cycles, \\ STT-RAM Write: 10 Cycles, \\ PCM Read: 35 Cycles, and \\ PCM Write: 100 Cycles \end{tabular} \\ \hline
\end{tabular}
}
\end{table}

\begin{table}[htb]
\centering
\caption{Nvsim parameters of SRAM, STT-RAM Caches, PCM memory (350K, 22nm)}
\label{tab2}

\begin{tabular}{|l|l|l|l|l|}
\hline
\textbf{Parameter} & \textbf{\begin{tabular}[c]{@{}l@{}}16KB\\ SRAM\end{tabular}} & \textbf{\begin{tabular}[c]{@{}l@{}}16KB\\ STT-RAM\end{tabular}} & \textbf{\begin{tabular}[c]{@{}l@{}}128KB\\ STT-RAM\end{tabular}} & \textbf{\begin{tabular}[c]{@{}l@{}}128MB\\ PCM\end{tabular}} \\ \hline
\textbf{Read Latency} & 0.792 ns & 1.994 ns & 1.861 ns & 204.584ns \\ \hline
\textbf{Read Energy} & 0.006 nJ & 0.081 nJ & 0.123 nJ & 1.553 nJ \\ \hline
\multirow{2}{*}{\textbf{Write Latency}} & \multirow{2}{*}{0.772 ns} & \multirow{2}{*}{10.520 ns} & \multirow{2}{*}{10.446 ns} & \begin{tabular}[c]{@{}l@{}}RESET - \\ 134.923 ns\end{tabular} \\ \cline{5-5} 
&  &  &  & \begin{tabular}[c]{@{}l@{}}SET - \\ 264.954 ns\end{tabular} \\ \hline
\multirow{2}{*}{\textbf{Write Energy}} & \multirow{2}{*}{0.002 nJ} & \multirow{2}{*}{0.217 nJ} & \multirow{2}{*}{0.542 nJ} & \begin{tabular}[c]{@{}l@{}}RESET - \\ 6.946 nJ\end{tabular} \\ \cline{5-5} 
 &  &  &  & \begin{tabular}[c]{@{}l@{}}SET - \\ 6.927 nJ\end{tabular} \\ \hline
\end{tabular}
\end{table}

\subsection{Baseline Architecture} \label{p51}

\begin{table}[htbp]
\centering
\caption{Overview of Baseline Architectures Configurations}
\label{ba}
\resizebox{\columnwidth}{!}{%
\begin{tabular}{|l|l|l|}
\hline
\textbf{Architecture} & \textbf{Memory} & \textbf{Policy} \\ \hline
\multirow{3}{*}{\textbf{Baseline-1}} & L1: SRAM (32 KB) & Write-through \\ \cline{2-3} 
& LLC: STT-RAM (256 KB) & Write back \\ \cline{2-3} 
& Main Memory: PCM (128 MB) & - \\ \hline
\multirow{3}{*}{\textbf{Baseline-2}} & L1: SRAM (32 KB) & Write back \\ \cline{2-3} 
& LLC: STT-RAM (256 KB) & Write back \\ \cline{2-3} 
& Main Memory: PCM (128 MB) &  \\ \hline
\multirow{3}{*}{\textbf{Baseline-3}} & L1: SRAM (4 KB) & Write back \\ \cline{2-3} 
& LLC: STT-RAM (256 KB) & Write back \\ \cline{2-3} 
& Main Memory: PCM (128 MB) &  \\ \hline
\end{tabular}
}
\end{table}

\begin{table*}[htpb]
\centering
\caption{Overview of the Different Possibilities for the Proposed Architecture that are Used for the Comparisons}
\label{base}
\begin{tabular}{|llllll|}
\hline
\multicolumn{1}{|l|}{\textbf{Proposed Architecture}} & \multicolumn{1}{l|}{\textbf{\begin{tabular}[c]{@{}l@{}} Proposed Base Architecture \\ (L1: SRAM, LLC-STT-RAM, Memory: PCM)\end{tabular}}} & \multicolumn{1}{l|}{\textbf{\begin{tabular}[c]{@{}l@{}} Proposed Techniques\\ (DBT, WBQ)\end{tabular}}} & \multicolumn{1}{l|}{\textbf{LFW}} & \multicolumn{1}{l|}{\textbf{LRW}} & \multicolumn{1}{l|}{\textbf{BR}} \\ \hline
\multicolumn{1}{|l|}{\textbf{Proposed with LRW}}        & \multicolumn{1}{l|}{\cmark}      & \multicolumn{1}{l|}{\cmark}              & \multicolumn{1}{l|}{\xmark}              & \multicolumn{1}{l|}{\cmark}               & \multicolumn{1}{l|}{\cmark}                               \\ \hline

\multicolumn{1}{|l|}{\textbf{Proposed without BR }}   & \multicolumn{1}{l|}{\cmark}         & \multicolumn{1}{l|}{\cmark}              & \multicolumn{1}{l|}{\cmark}              & \multicolumn{1}{l|}{\xmark}               & \multicolumn{1}{l|}{\xmark}                                  \\ \hline
\multicolumn{1}{|l|}{\textbf{Proposed without BR and with LRW}}   & \multicolumn{1}{l|}{\cmark}         & \multicolumn{1}{l|}{\cmark}              & \multicolumn{1}{l|}{\xmark}              & \multicolumn{1}{l|}{\cmark}               & \multicolumn{1}{l|}{\xmark}                                \\ \hline
\multicolumn{1}{|l|}{\textbf{Proposed Architecture}}       & \multicolumn{1}{l|}{\cmark}      & \multicolumn{1}{l|}{\cmark}              & \multicolumn{1}{l|}{\cmark}              & \multicolumn{1}{l|}{\xmark}               & \multicolumn{1}{l|}{\cmark}                                   \\ \hline
\multicolumn{6}{|l|}{\cmark - Supported ,   \xmark - Not Supported}                                                                                                                                                                                                         \\ \hline

\end{tabular}%
% }
\end{table*}

We modeled three baseline architectures to compare with the proposed architecture as shown in table \ref{ba}.

\begin{itemize}
    \item At L1, Baseline-1 uses a write-through policy. As a result, L1 contains no dirty blocks. Baseline-1 uses the least amount of backup energy during a power failure. 
    \item At L1, Baseline-2 uses a write-back policy. As a result, the number of LLC writes decreases. Baseline-2 improves system performance over baseline-1.
    \item Baseline-3 has a 4 KB L1 cache and the same LLC and main memory sizes as in baseline-2. Baseline-3 assists in determining whether using small-size volatile memory at L1 improves performance during a stable power supply. During a power failure, our proposed backup contents are the same size as baseline-3.
\end{itemize}

Under a stable power supply, we compared baselines-1, 2, and 3 to the proposed architecture. We compared baseline-2 to the proposed architecture during frequent power failures. Proposed policies such as DBT, WBQ, and replacement policies are not included in the baselines-1,2, and 3 architectures.

\subsection{Results} \label{p52}
The proposed architecture is evaluated in this section under stable power and power failures. The proposed architecture is compared with two baseline architectures.

\subsubsection{\textbf{Proposed Architecture Under Stable Power Supply}} \label{p521}

\begin{figure}[htb]
   \includegraphics[width= 1\linewidth]{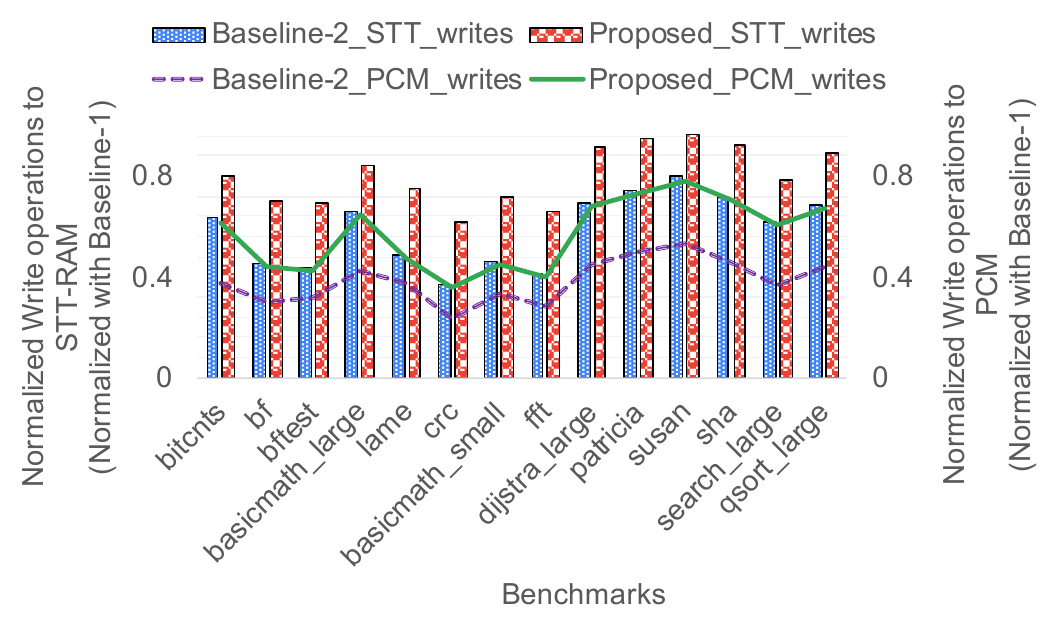}
   \centering
 \caption{Write operations for STT-RAM, PCM under Stable Power.}
 \label{fig2a}
 \end{figure}

\begin{figure}[htb]
  \includegraphics[width= 1\linewidth]{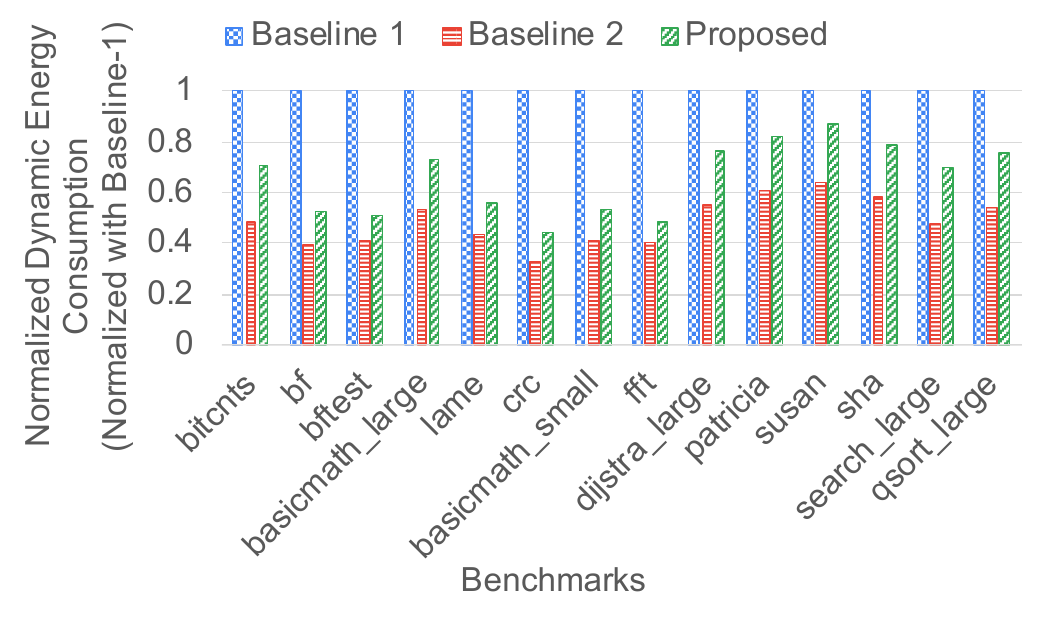}
  \centering
 \caption{Comparisons between Proposed and Baseline Architectures for Dynamic Energy Consumption under Stable Power.}
 \label{fig2c}
 \end{figure}

\begin{figure}[htb]
   \includegraphics[width= 1\linewidth]{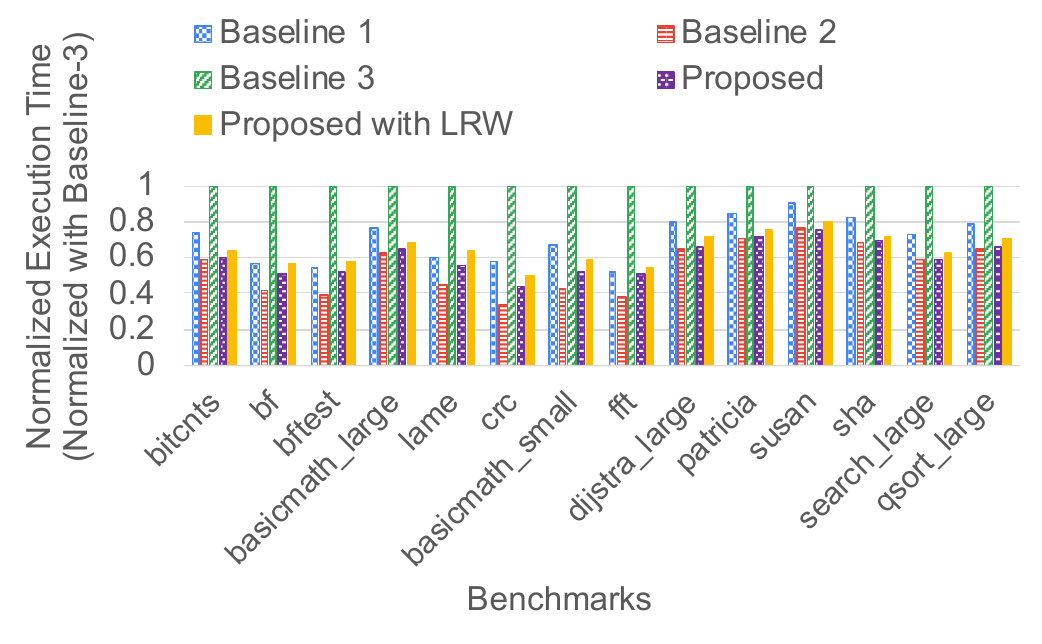}
   \centering
 \caption{Comparisons between Proposed and Baseline Architectures for Execution Time under Stable Power.}
 \label{fig2b}
 \end{figure}

In order to reduce the system's energy consumption, we first need to reduce the number of writes to both STT-RAM and PCM. We performed experiments to compare the number of writes for NVM in baseline and proposed architectures. All values shown in figures \ref{fig2a} and \ref{fig2c} are normalized with the baseline-1 architecture. All values shown in figure \ref{fig2b} are normalized with the baseline-3 architecture. Baseline-2 gets fewer writes than the proposed architecture at LLC and PCM by 14.11\% and 7.84\%, as shown in figure \ref{fig2a}. Compared to baseline-1, the proposed gets 18.97\% fewer writes at the LLC and 10.66\% at the PCM, as shown in figure \ref{fig2a}.

As shown in figure \ref{fig2c}, the proposed architecture consumes 17.56\% less energy than baseline-1 architecture and 4.93\% more energy than baseline-2 architecture. Under a stable power supply scenario, the proposed architecture consumes less energy than baseline-1 and more energy than baseline-2. 

During a power failure, the proposed architecture backup only the contents of DBT and WBQ to LLC. The combined size of DBT and WBQ is approximately less or equal to 4 KB. As a result, we compared the proposed architecture to baseline-3 to see how it performs in normal operations. As shown in figure \ref{fig2b}, architecture-3 performs poorly compared to the proposed and baseline architectures. Compared to baseline-3, the proposed architecture takes 38.79\% less execution time during regular operation. Thus, using a small L1 size cache doesn't improve performance during a stable power supply.

As shown in figure \ref{fig2b}, we compared architectures with LRW and LFW policies.  As shown in table \ref{base}, the proposed architecture uses the LFW replacement policy, and the proposed using LRW architecture uses the LRW replacement policy instead of LFW. As shown in figure \ref{fig2b}, the proposed architecture takes 13.11\% less execution time than baseline-1 architecture and 5.10\% more execution time than baseline-2 architecture. The proposed architecture with the LFW policy performs better than the architecture with the LRW policy.

\begin{figure}[htb]
  \includegraphics[width= 1\linewidth]{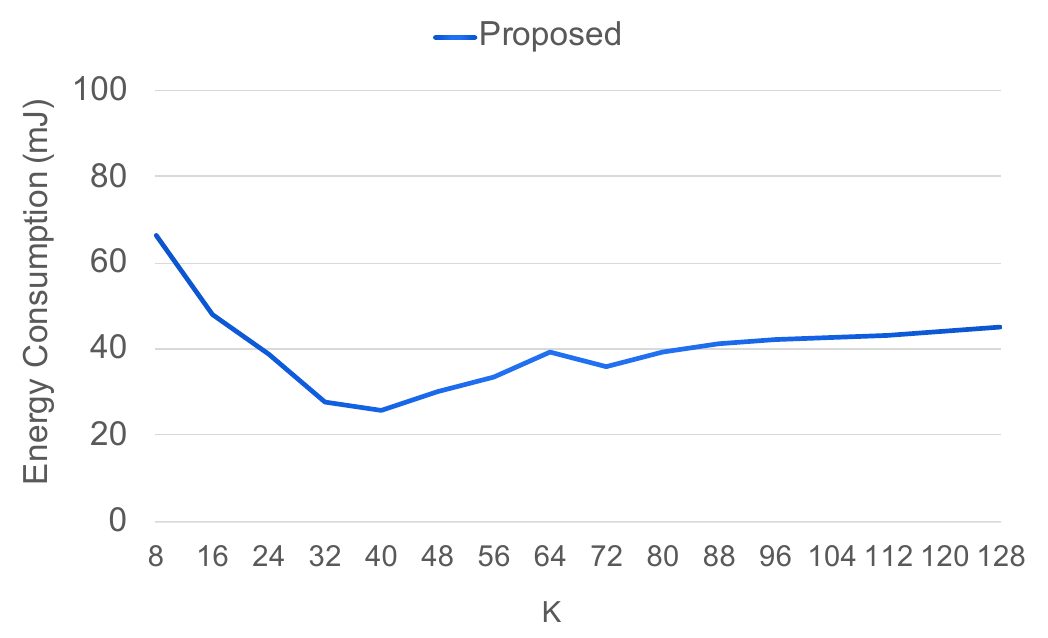}
 \caption{Energy Consumption for Different K Values under Stable Power.}
 \label{fig2e}
 \end{figure}

\begin{figure}[htb]
  \includegraphics[width= 1\linewidth]{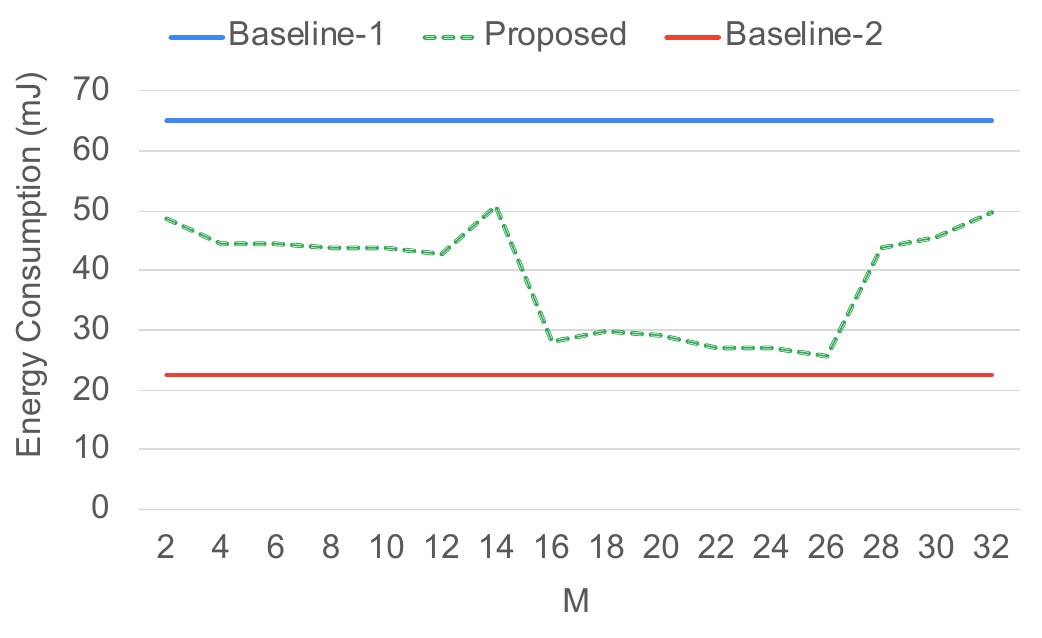}
 \caption{Energy Consumption for Different M, N values under Stable Power. }
 \label{fig2d11}
 \end{figure}

We performed experiments to determine K, M, and N values as shown in the figures \ref{fig2e}, \ref{fig2d11}. We performed the experiments for various $K$ values, as shown in figure \ref{fig2e}. We performed experiments with various $K$ values from K=8 to K=128. For all experiments shown in figure \ref{fig2e}, we assumed (M, N) to be equal. For example, if K=16, M=N=8. As shown in figure \ref{fig2e}, increasing the $K$ value consumes higher energy values. We assume capacitor energy as input; thus, $K$ is also input to our design based on equation \ref{eq4}. We analyzed various $K$ values because different sizes of capacitors are available in the market. The size of $M$ entirely decides the LRW field's size. Figure \ref{fig2d11} already show the energy consumption values for various $M$ values. Based on the $M$ value, we can set the LRW field's size. If we look at the figure \ref{fig2e}, we observe that the system uses less energy up until K=42, then gradually consumes more energy as K increases.

As shown in figure \ref{fig2d11}, we used K as 32 in these experiments. Within K=32, we experimented on various (M, N) pairs. In figure \ref{fig2d11}, when M=2, N becomes 14 (16-M), and when M=12, N becomes 4. We performed experiments with various possible (M, N) pairs such as (6, 10), (16, 16), (24, 8), and (32, 0). We observe that (26, 6) utilizes less energy than the other pairs. As a result, for K=32, we used (M, N) as (26, 6) throughout this paper. In the same way, we experimented to identify the best (M, N) pair for K=64. We observe that (54, 10) uses less energy than all other pairs for K=64.

 We performed experiments on the same benchmark suite used in section \ref{exp1} to determine the K, M, and N values. We selected K, M, and N values based on the average values. The optimal K, M, and N values vary depending on the application behavior and the system configuration. We also observed that for the selected K, M, and N values, the majority of benchmarks (11 out of 14) outperform others. The variation between other K, M, and N values and selected values is negligible for the other three benchmarks. As a result, we selected K, M, and N values after considering all benchmarks. We have shown the overall average values in figures \ref{fig2e} and \ref{fig2d11}. Table \ref{tab1} lists the selected K, M, and N values.

As shown in figure \ref{fig2d2}, our proposed architecture has four design alternatives. Using NVM at both L1 and LLC in architecture-1 gives us the third alternative. Adding BR to architecture-2 gives us the fourth alternative. 

\begin{figure}[htb]
  \includegraphics[width= 1\linewidth]{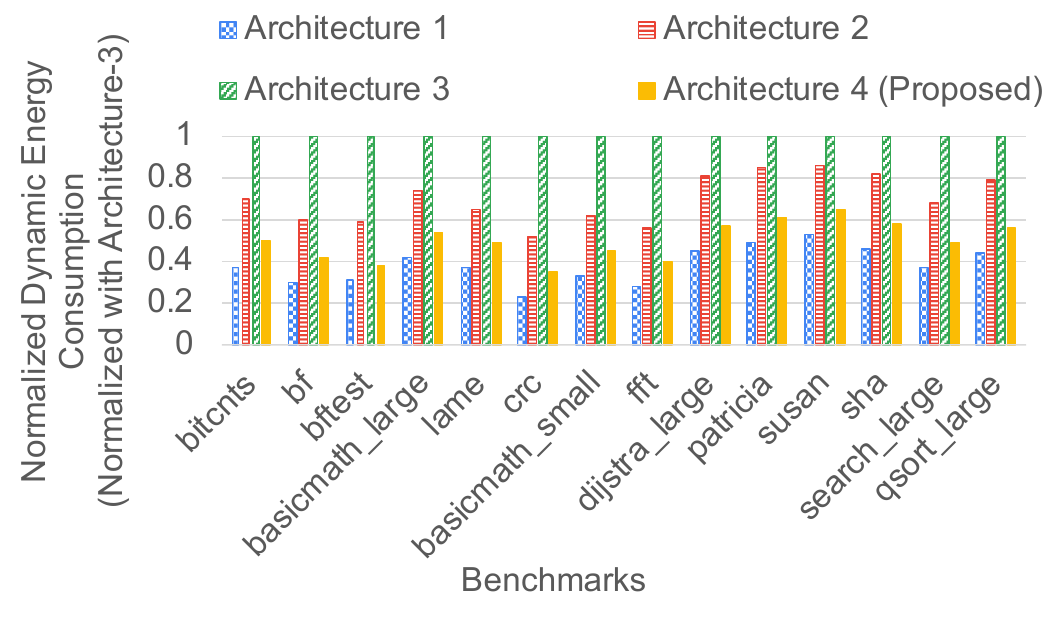}
 \caption{ Comparisons between Proposed Architecture and Unified NVM-based Architectures for Dynamic Energy Consumption under Stable Power.}
 \label{cp110}
 \end{figure}

We performed experiments with the unified NVM architecture, i.e., architecture-3. We compared unified NVM architecture to the proposed architecture to determine how good or bad architecture-3 will perform under a stable power supply. We compared architecture-3 to architectures-1 and 2 for dynamic energy consumption, as shown in figure \ref{cp110}. Under a stable power supply, architecture-1 outperforms architectures-2, 3, and the proposed architecture. This advantage is due to the use of volatile memory at L1 and LLC in architecture-1.

 We observed that NVM receives more read/write accesses when the entire memory at L1 is NVM-based. Compared to the proposed architecture, architecture-3 consumes 43.25\% more energy and has a performance overhead of 38.99\% under a stable power supply. This analysis motivated us to compare the proposed architecture with one that uses NVM-based memory for only DBT and WBQ at the L1 cache. The proposed architecture consumes 19.41\% less energy than this design.

\begin{figure}[htb]
  \includegraphics[width= 1\linewidth]{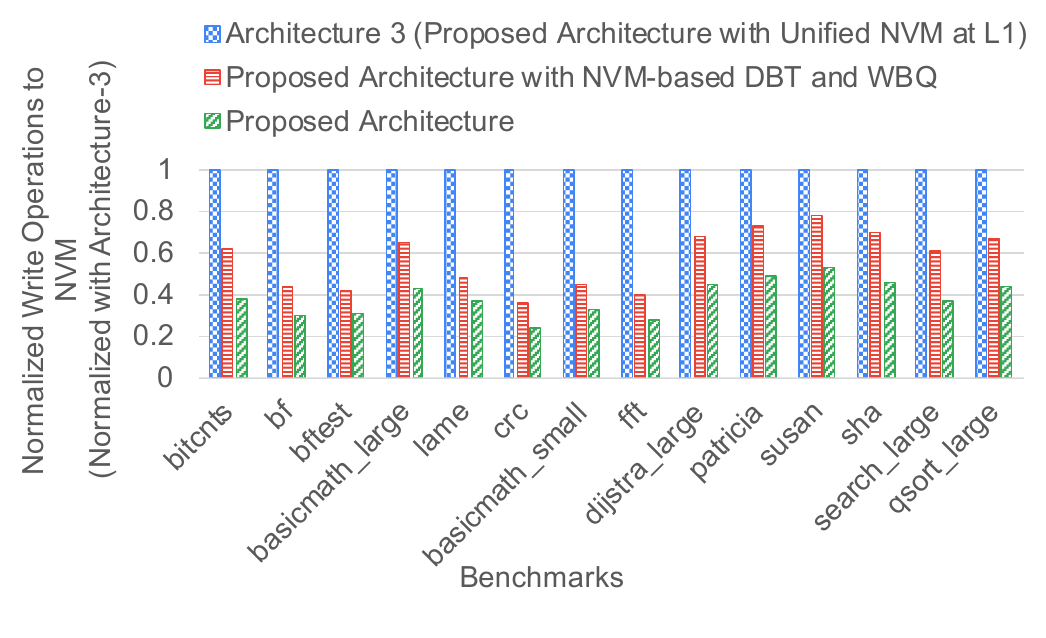}
 \caption{ Comparisons between Proposed Architecture and Unified NVM-based Architectures for NVM Write operations under Stable Power.}
 \label{cp1100}
 \end{figure}

 We performed experiments to compare architecture-3 with the architecture that uses NVM-based DBT and WBQ for the number of NVM writes, as shown in figure \ref{cp1100}. Under a stable power, NVM-based DBT and WBQ design consumes 37.17\% lesser writes than architecture-3 and 21.79\% more writes than the proposed architecture. As a result, unified NVM architecture is not a good choice to use for regular operations.

 \textbf{Analysis with Different Checkpointing Approaches:} The proposed backup/restore strategy looks similar to checkpointing. We performed experiments with a stable power supply to compare the proposed backup/restore strategy to existing checkpointing approaches. We used three different checkpointing methods for these experiments.

\begin{figure}[htb]
  \includegraphics[width= 1\linewidth]{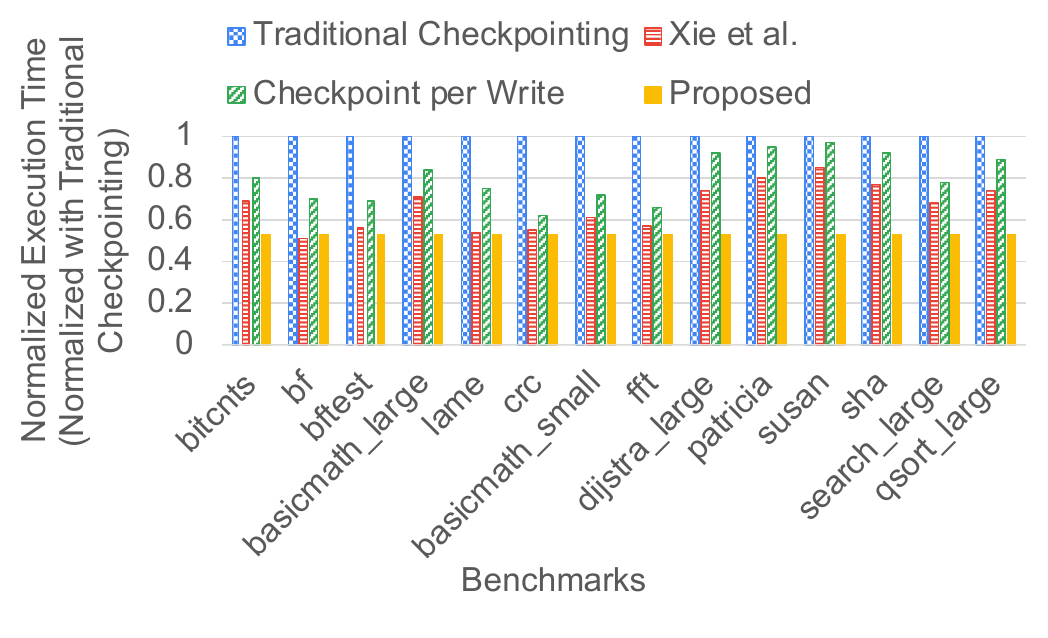}
 \caption{ Comparisons between Proposed Backup and Different Existing Checkpointing Techniques for Execution Time under Stable Power.}
 \label{cp1}
 \end{figure}

\begin{figure}[htbp]
  \includegraphics[width= 1\linewidth]{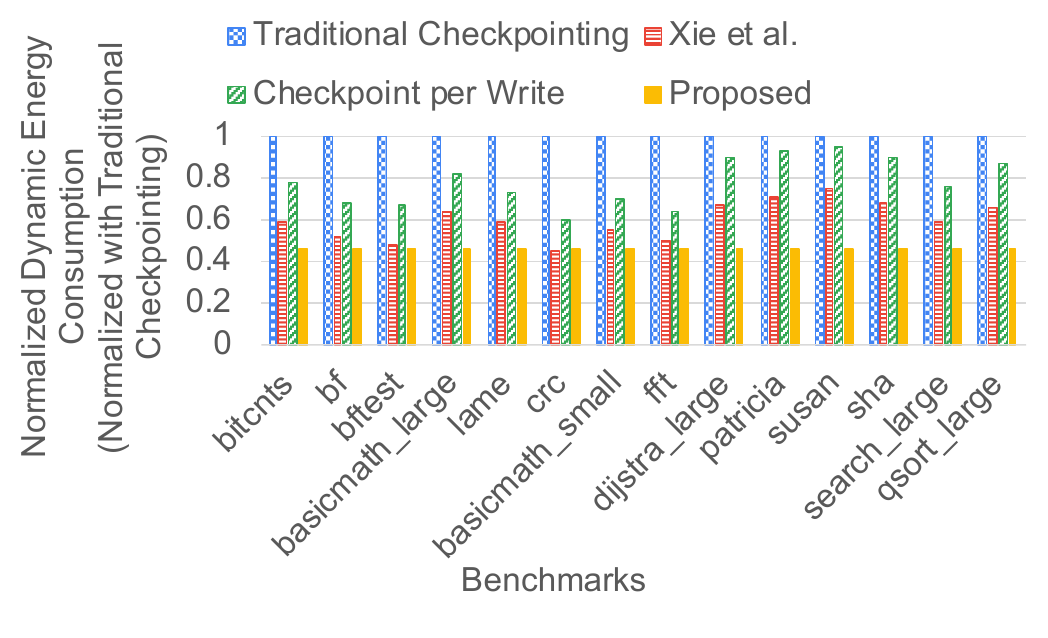}
 \caption{ Comparisons between Proposed Backup and Different Existing Checkpointing Techniques for Energy Consumption under Stable Power. }
 \label{cp2}
 \end{figure}

 First, we designed a traditional checkpointing technique, introducing a safe point at every 4 million instructions. For every 4 million instructions, we backup the system state to NVM. We restore from the main memory at each safe point and continue with the application's execution. Our proposed architecture outperforms traditional checkpointing because backup only occurs during power failure. All values shown in figures \ref{cp1} and \ref{cp2} are normalized with the traditional checkpointing technique. As shown in figures \ref{cp1} and \ref{cp2}, the proposed architecture reduces performance overhead and energy consumption by 41.27\% and 37.95\%. 

 Second, we used a checkpointing method proposed by Xie et al. \cite{xie}. Xie et al. identified the important volatile blocks that needed to be backed up during a power failure using STT-RAM-based counters (DBCounter and MCounter). Xie et al. used the LRU policy for replacement in NVM-based caches and volatile caches. Updating and accessing these counters is similar to NVM writes and reads, which use more energy and slow down the system. Because Xie et al. performs backup during a power failure, we compared performance and energy consumption based on total NVM reads/writes. As shown in figures \ref{cp1} and \ref{cp2}, the proposed architecture reduces performance overhead and energy by 19.03\% and 14.72\%.

 Third, we implemented a checkpointing procedure that creates a checkpoint whenever the system state changes. For every write request, we increment the write counter (WC). Once WC reaches the threshold, the checkpoint procedure is triggered, size of the WC size is the same as the proposed system configuration. This checkpointing technique increases NVM reads/writes, degrades system performance, and consumes more energy. As shown in figures \ref{cp1} and \ref{cp2}, the proposed architecture reduces performance overhead and energy consumption by 39.11\% and 33.10\%. 

 The traditional checkpointing technique is both the standard and the worst-case scenario for these intermittently powered systems. If we want to add another level of filtering checkpoints, we can add a counter per cache block, and if any cache block reaches the defined threshold, the checkpoint procedure is triggered. This checkpointing technique performs better than traditional checkpointing. However, the second checkpointing technique increases the number of NVM accesses. Instead of checkpointing at every safe point or for every $>WC$ case, we can add another level for reducing checkpoints by placing a checkpoint only during a power failure. Xie et al. proposed a checkpointing policy that only backups the dirty blocks selected during a power failure. Xie et al. outperform the other two checkpointing policies. However, the proposed backup strategy reduces performance overhead and energy consumption compared to Xie et al., as shown in figures \ref{cp1} and \ref{cp2}.

\subsubsection{\textbf{Proposed Architecture Under Unstable Power}}

We simulate frequent power failures by assuming that a power failure occurs for every 2 million instructions. The open-source version of the Gem5 core does not model an intermittent power supply processor. By introducing interrupts, we modified gem5 to support intermittent power supply processors. So, for every 2 million instructions, there is an interrupt, which the processor model admits as a power failure. The Gem5 simulator is used to run all of the experiments for one billion instructions. We assumed a power failure occurs for every 2 million instructions because, on average, 2 million instructions take approximately 25-30 ms of time to execute. In another way, there is a power interruption every 30 ms, so these power failures are not as frequent as they would be in real life. Therefore, the results are rather conservative.

Considering energy harvesting sources, such as piezoelectric and vibration-based sources, they extract significantly less energy from their surroundings. In such cases, the capacitor is unable to store sufficient energy, leading to frequent power failures. As a consequence, our proposed architecture is capable of handling these worst-case scenarios. However, existing work by Xie et al. \cite{xie} made similar assumptions that almost every power failure occurs every 200 and 500 ms.

\begin{figure}[htb]
\centering
   \includegraphics[width= 1\linewidth]{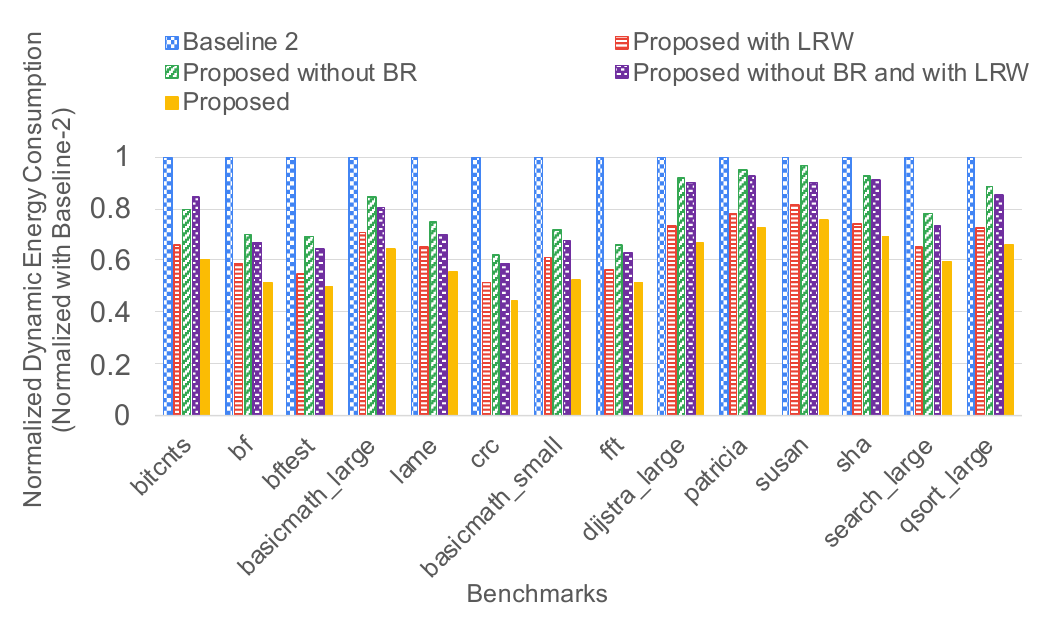}
 \caption{ Comparisons between Proposed and Baseline Architectures for Energy Consumption under Unstable Power.}
 \label{fig3a}
 \end{figure}

\begin{figure}[htb]
\centering
   \includegraphics[width= 1\linewidth]{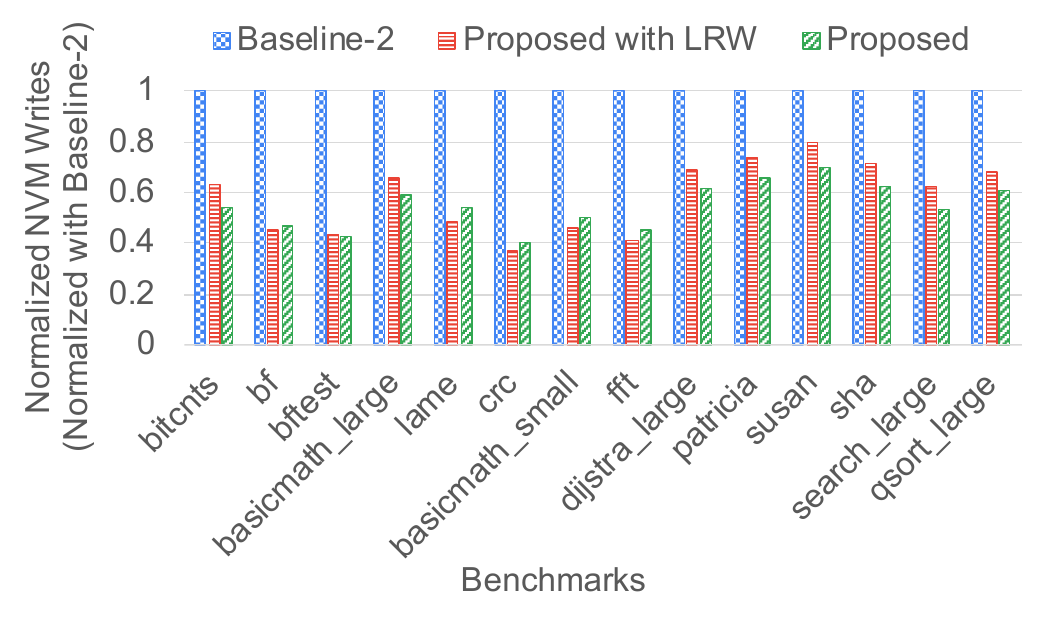}
 \caption{ Comparisons between LRW and LFW Replacement Policies with Proposed and Baseline Architectures under Unstable Power.}
 \label{fig3b}
 \end{figure}

All values shown in figures \ref{fig3a}, \ref{fig3b} are normalized with baseline-2 architecture. The proposed architecture consumes 20.94\% less energy than the baseline-2 architecture, as shown in figure \ref{fig3a}. We evaluated the proposed architecture with both replacement policies. The architecture that uses LFW performs better than LRW, as shown in figure \ref{fig3b}.

We performed a series of experiments to analyze the backup energy and the effect of PCM on our proposed architecture. 

\begin{figure}[htb]
  \includegraphics[width= 1\linewidth]{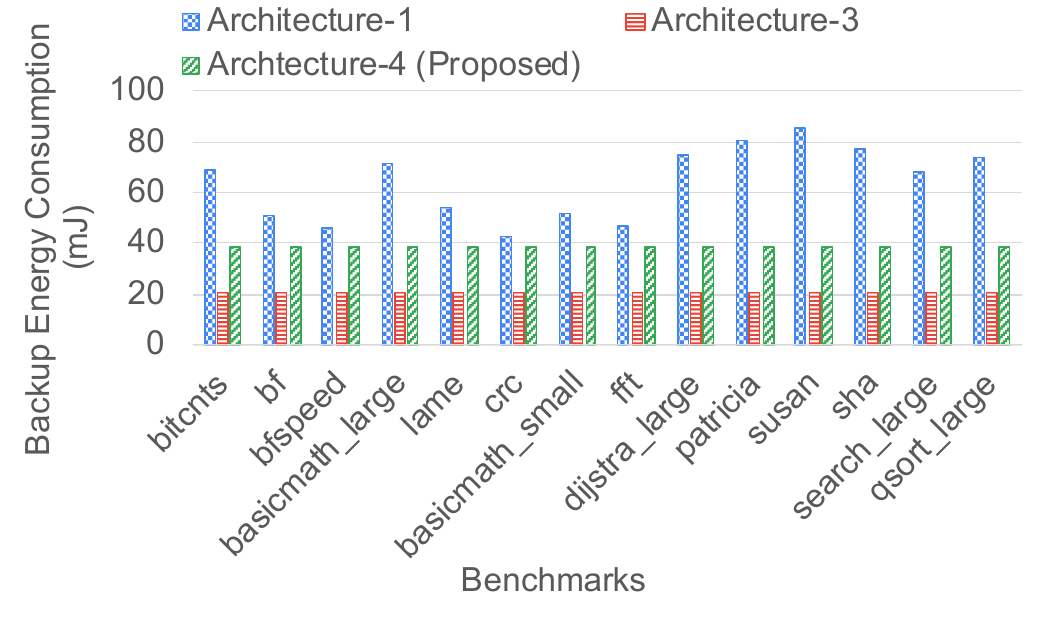}
 \caption{ Comparisons between Architecture-1, 3, and 4 for Backup Energy Consumption under Unstable Power.}
 \label{fig2d1}
 \end{figure}

During frequent power failures, architecture-3 outperforms architecture-1 and the proposed architecture for backup energy consumption. Compared with architecture-1, architecture-4 consumes 35.57\% less energy, as shown in figure \ref{fig2d1}. Because we need to backup the entire L1 and LLC to PCM, architecture-1 requires more backup energy than architecture-3. Due to the unified NVM architecture, architecture-3 only needs to backup volatile register contents. Architecture-4 requires constant backup energy because we only need to backup volatile register contents and $K$ blocks to BR during a power failure. Thus, architecture-4 consumes more energy than architecture-3. We have shown the required backup energy for architecture-1 and 4 in equation \ref{mot2}. Architectures-3 and 4 use constant energy to backup volatile contents.

\begin{figure}[htb]
  \includegraphics[width= 1.0\linewidth]{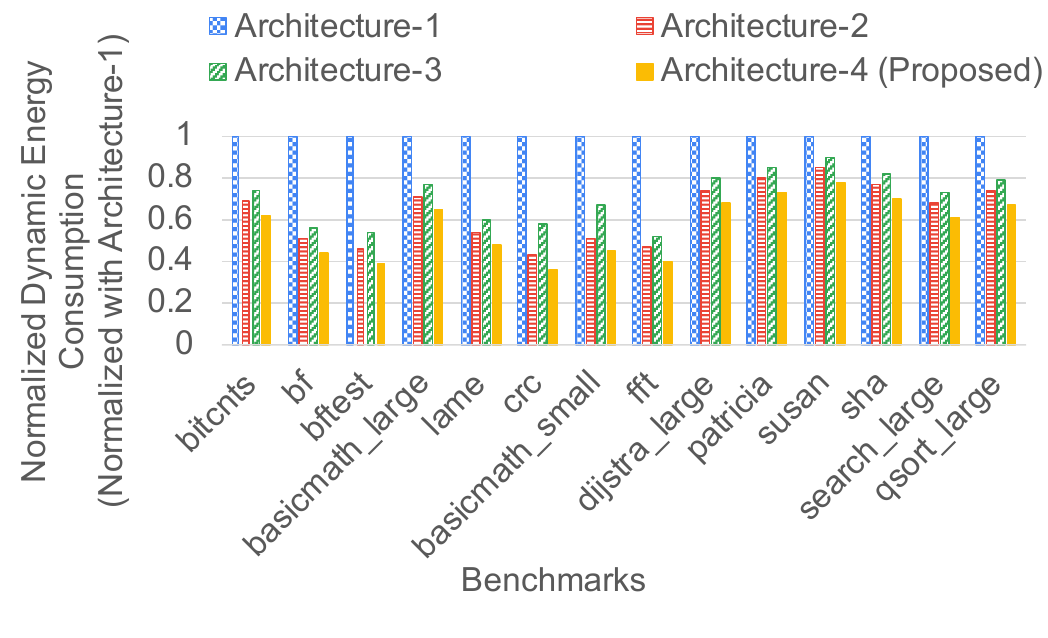}
 \caption{Comparisons between Architectures-1, 2, 3, and 4 for Overall Dynamic Energy Consumption under Intermittent Power Supply.}
 \label{fig2d21}
 \end{figure}

We want to examine whether all these benefits are because of PCM at the main-memory level. So, we compare the energy consumption for four architectures shown in figure \ref{fig2d21}. As shown in figure \ref{fig2d21}, architecture-1 consumes more energy than the other three architectures because the number of writes to PCM is more in architecture-1. The only difference between architecture-2 and 4 is that architecture-4 is BR-enabled.

The performance of BR-enabled architecture is more effective than that of non-BR-enabled architecture, i.e., architecture-2. With a BR at LLC, we can directly place those $K$ blocks in BR and quickly restore the contents of the L1 cache. When we remove BR at LLC, we must first update in LLC. If LLC is full, we need to replace some blocks at LLC to make space for the L1 dirty blocks. We use the LRU replacement policy at LLC, which increases write to PCM compared to the BR-enabled architecture. We assume our system has a fixed-energy capacitor that can only backup the $K$ blocks and the register file. When we remove BR from LLC, the capacitor no longer supports safe backup because we have to lose $K$ or $< K$ blocks (either from L1 or LLC). As a result, we either end up with the wrong results or have to restart the application. As illustrated in figure \ref{fig3a}, BR-enabled architecture consumes less energy than the architecture without BR at LLC. This benefit is because of the increased number of writes to PCM. The proposed architecture consumes less energy than the baseline-2 architecture because frequent power failures increase reads/writes to NVM, as shown in figure \ref{fig3a}. 

Compared with architecture-1, architecture-2 consumes 19.02\% less energy, and architecture-4 consumes 32.64\% less energy. Figure  \ref{fig2d21} shows that architecture-4 is better than the other three architectures. Therefore, not all of these improvements are solely linked to PCM since our proposed policies also help to achieve better performance and energy.

We performed experiments with the unified NVM architecture, i.e., architecture-3. All values shown in figure \ref{fig2d21} are normalized with architecture-1. Under frequent power failures, architecture-3 outperforms architecture 1, but not architecture-2 and the proposed architecture. This advantage for architecture-3 is due to the usage of NVM at L1 and LLC. 

Architecture-3 consumes 23.01\% more energy than the proposed architecture, i.e., architecture-4, as shown in figure \ref{fig2d21}. This benefit for the proposed architecture is because the energy required for backing up the $K$ blocks is less than that of the energy required for architecture-3 during regular operations. Figures \ref{cp110} and \ref{cp1100} show that architecture-3 consumes more energy and attracts more NVM reads and writes during regular operations. Architecture-3 is not suitable for intermittent power supply scenarios due to these overheads during normal operations. The unified NVM architecture requires additional procedures to make the system more energy efficient.

\textbf{Analysis with Different Checkpointing Approaches:} The proposed backup/restore strategy looks similar to checkpointing, a widely used approach to save the system state during a power failure. We performed experiments to compare the proposed backup/restore strategy with the existing checkpointing approaches under frequent power failures. For these experiments, we used four different checkpointing methods.

\begin{figure}[htbp]
  \includegraphics[width= 1\linewidth]{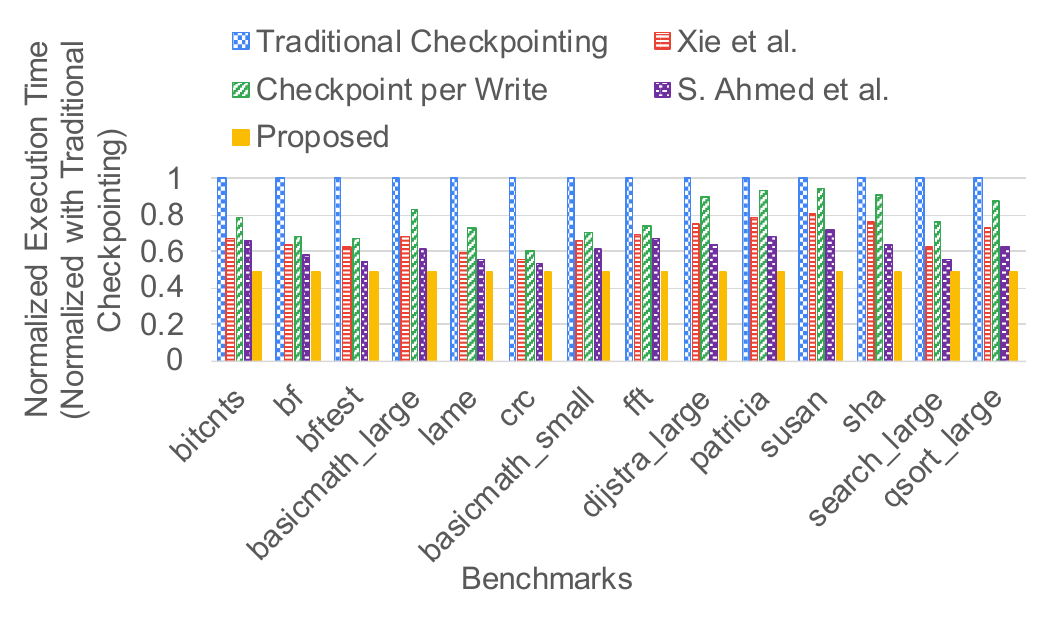}
 \caption{Comparisons between Proposed Backup and Different Existing Checkpointing Techniques for Execution Time under Unstable Power.}
 \label{cp11}
 \end{figure}

\begin{figure}[htb]
  \includegraphics[width= 1\linewidth]{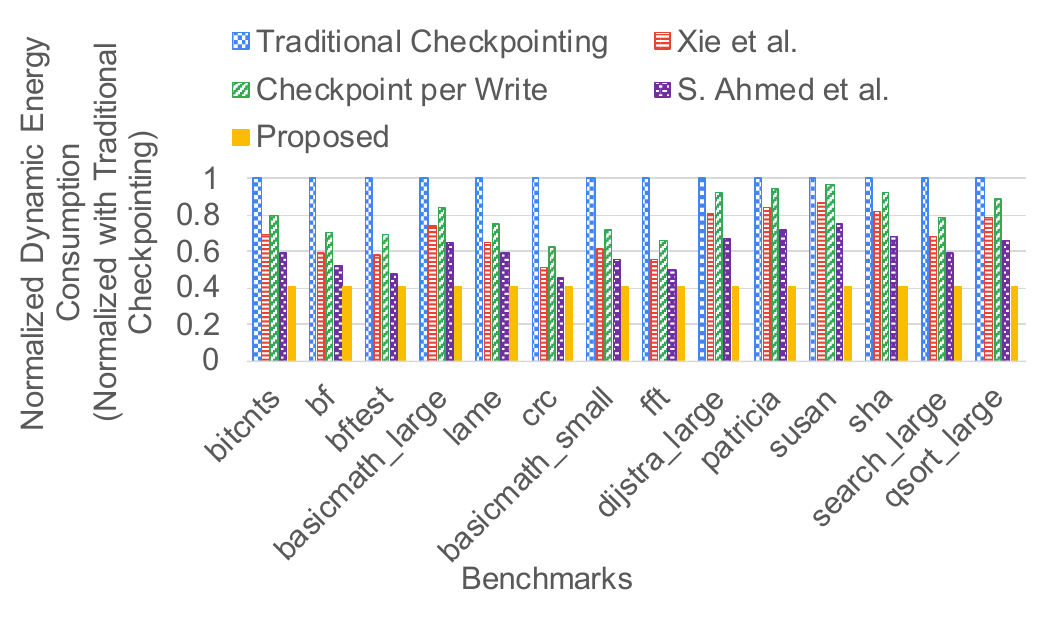}
 \caption{Comparisons between Proposed Backup and Different Existing Checkpointing Techniques for Energy Consumption under Unstable Power. }
 \label{cp21}
 \end{figure}

First, we designed a traditional checkpointing technique. Section \ref{p521} has discussed the checkpointing technique's implementation details. All values shown in figures \ref{cp11} and \ref{cp21} are normalized with the traditional checkpointing technique. For example, suppose an unexpected power failure occurs at the $7th$ million instruction. We re-execute the application from the $4th$ million instruction because it is the closest safe point. These unnecessary executions increase NVM writes/reads, consume more energy, and degrade system performance. The proposed architecture reduces performance overhead and energy consumption by 48.70\% and 40.19\%, as shown in figures \ref{cp11} and \ref{cp21}.

Second, we used a checkpointing method proposed by Xie et al. \cite{xie}. The implementation details of the Xie et al. checkpointing technique are discussed in section \ref{p521}. The proposed architecture reduces performance overhead and energy consumption by 27.99\% and 20.07\%, as shown in figures \ref{cp11} and \ref{cp21}. This advantage is due to the author's selection of an LRU-based replacement policy for updating the counters, which we observed as insufficient to reduce the number of writes to NVM. Another reason is that Xie et al. did not backup the entire dirty block to NVM, which requires the application to be re-executed, which consumes more energy during frequent power failures.

Third, we implemented a checkpointing procedure that creates a checkpoint whenever the system state changes. Section \ref{p521} has discussed the checkpointing technique's implementation details. The proposed architecture reduces performance overhead and energy consumption by 38.71\% and 32.13\%, as shown in figures \ref{cp11} and \ref{cp21}. During frequent power failures, backup/restore sizes increase in this technique, potentially causing NVM reads and writes to increase, which degrades system performance and consumes more energy.

 Fourth, we implemented a checkpointing procedure that initiates a backup procedure whenever the system state changes. During a power failure, the checkpoint procedure is triggered, and we compare the previous checkpoint data to the new checkpoint data, block by block. Only the volatile contents that differ from the previous checkpoint data are backed up. Copy-by-Change checkpointing procedures are proposed by S. Ahmed et al. \cite{check2,check3}. The proposed architecture reduces performance overhead and energy consumption by 21.93\% and 16.55\%, as shown in figures \ref{cp11} and \ref{cp21}.

 We already discussed the first three checkpointing policies in section \ref{p521}. In order to add another level to Xie et al. for reducing the checkpointing overhead. Instead of checkpointing all cache blocks, S. Ahmed et al. policy checks and compares the most recently checkpointed data. When compared to Xie et al., this policy reduces NVM accesses; thus, the S. Ahmed et al. checkpointing policy outperforms the Xie et al. policy. However, accessing and comparing NVM-based previous checkpointed data also induces reads/writes to NVM. Figures \ref{cp11} and \ref{cp21} show that the proposed backup strategy performs better than the existing checkpointing techniques.

 We compared the average memory access time (AMAT) for two architectures. One, the architecture is without any proposed data structures and replacement policies, i.e., the baseline-2 architecture. Second, the architecture with proposed data structures and replacement policies. We performed these experiments to analyze the proposed architecture to see how the proposed data structures and replacement policy impacted system performance. AMAT analysis allows us to determine which of the two architectures is faster during frequent power failures. The proposed architecture outperforms the baseline-2 architecture by 19.61\% during frequent power failures. This advantage is due to DBT and WBQ; most L1 requests result in hits, which reduces performance overhead. Another significant advantage of the proposed architecture is the explored replacement policies. During a stable power supply, the proposed replacement policy helps in saving important blocks (write-intensive blocks) at the L1 cache rather than evicting them. Thus, incorporating DBT and WBQ improves the proposed architecture's performance under intermittent power.

\subsection{Analysis for Multi-Cache Levels and Multi-Core Designs}

 The proposed architecture consists of an SRAM-based L1 cache and an STT-RAM-based LLC. We found two possibilities if we want to add one or two levels of cache to the proposed architecture.

 First, suppose we introduce one or two levels of STT-RAM-based caches. In that case, our proposed architecture has no impact or complications with this design because we proposed a backup strategy that can only track and backup $K$ dirty blocks from the L1/L2/L3 caches. Everything is the same as in the proposed architecture during a stable power supply. We don't need a backup of L2/L3 during frequent power failures because these are already NVM-based caches. According to our proposed backup policy, we must back up the contents of registers and the L1 dirty contents to LLC. We must track the number of dirty block contents in the L1 cache. This appears to be similar to our proposed architecture and techniques. As a result, our proposed architecture does not require additional methods and data structures for the first possibility.

 Second, imagine we have one or two levels of SRAM-based caches. In that case, our proposed architecture makes a difference. This design is complicated because it is necessary to limit the total number of dirty blocks from all levels of SRAM-based caches, i.e., from L1/L2/L3 caches to $K$ at any point in time. This issue requires several changes to the proposed architecture for tracking and maintaining $K$ dirty data at each level. If the processor is multi-core, the problem becomes even more complex and introduces new complications, such as cache coherence issues. Selecting volatile data and backing it up to NVM requires significant energy and additional techniques during frequent power failures. These extra procedures add various overheads, such as size and performance. However, these complex systems are not required for embedded devices or applications.

 NVM-enabled microcontrollers do not contain a second-level cache. Texas Instruments (TI)-based NVPs, such as the MSP430FR6989 and MSP430F5529, don't have a cache and contain only main memory. The main memory of the MSP430FR6989 contains 2KB of SRAM and 128KB of FRAM, while the MSP430F5529 contains SRAM and flash. For these intermittently powered IoT systems to run embedded applications, we don't need a multi-core processor with complex cache hierarchies; instead, one/two levels of caches and a 1-core processor are sufficient. As a result, in this work, we did not examine the proposed architecture for higher levels of cache or multi-core processors.

 \begin{table*}[htp]
\centering
\caption{Comparison of Different Possibilities for a given Cache size(L1/LLC) as 32KB/256KB and Main memory size as 64MB}
\label{tabc2}

\begin{tabular}{|l|l|l|l|l|l|l|l|l|l|l|}
\hline
\textbf{\begin{tabular}[c]{@{}l@{}}Cache size\\  (L1/LLC)\end{tabular}} & \textbf{\begin{tabular}[c]{@{}l@{}}Main Memory \\ size\end{tabular}} & \textbf{K} & \textbf{M} & \textbf{N} & \textbf{WC} & \textbf{LRW/LFW} & \textbf{BR-enabled} & \textbf{\begin{tabular}[c]{@{}l@{}}Number of \\ power failures\end{tabular}} & \textbf{\begin{tabular}[c]{@{}l@{}}Energy \\ Gain (\%)\end{tabular}} & \textbf{Compared with} \\ \hline
\multirow{20}{*}{32KB/256KB}                                            & \multirow{20}{*}{64MB}                                                                                                          & 8          & 6          & 2          & 6           & LFW              & Yes                 & 500                                                                          & 13.60                                                               & Baseline-2             \\ \cline{3-11} 
                                                                        &                                                                      & 8          & 6          & 2          & 0           & LRW              & Yes                 & 500                                                                          & 12.35                                                               & Baseline-2             \\ \cline{3-11} 
                                                                        &                                                                      & 8          & 6          & 2          & 6           & LFW              & No                  & 500                                                                          & 7.15                                                                & Baseline-2             \\ \cline{3-11} 
                                                                        &                                                                                                                         & 8          & 6          & 2          & 6           & LFW              & Yes                 & 200                                                                          & 11.76                                                               & Baseline-2             \\ \cline{3-11} 
                                                                        &                                                                      & 8          & 6          & 2          & 6           & LFW              & No                  & 200                                                                          & 6.84                                                                & Baseline-2             \\ \cline{3-11} 
                                                                        &                                                                      & 8          & 6          & 2          & 6           & LFW              & Yes                 & 1000                                                                         & 14.32                                                               & Baseline-2             \\ \cline{3-11} 
                                                                        &                                                                      & 8          & 6          & 2          & 6           & LFW              & No                  & 1000                                                                         & 7.58                                                                & Baseline-2             \\ \cline{3-11} 
                                                                        &                                                             
                                                                                                                                           & 16         & 8          & 8          & 6           & LFW              & Yes                 & 500                                                                          & 15.37                                                               & Baseline-2             \\ \cline{3-11} 
                                                                        &                                                                      & 16         & 12         & 4          & 0           & LRW              & Yes                 & 500                                                                          & 15.56                                                               & Baseline-2             \\ \cline{3-11} 
                                                                        &                                                                      & 16         & 12         & 4          & 6           & LFW              & No                  & 500                                                                          & 8.230                                                                & Baseline-2             \\ \cline{3-11} 
                                                                        &                                                                                 & 16         & 12         & 4          & 6           & LFW              & Yes                 & 200                                                                          & 15.64                                                               & Baseline-2             \\ \cline{3-11} 
                                                                        &                                                                      & \textbf{16}         & \textbf{12}         & \textbf{4}          & \textbf{6}           & \textbf{LFW}              & \textbf{Yes}                 & \textbf{1000}                                                                         & \textbf{18.04}                                                               & \textbf{Baseline-2}             \\ \cline{3-11} 
                                                                        &                                                                      & 16         & 8          & 8          & 6           & LFW              & Yes                 & 200                                                                          & 14.45                                                               & Baseline-2             \\ \cline{3-11} 
                                                                        &                                                                      & 16         & 8          & 8          & 6           & LFW              & No                  & 1000                                                                         & 9.43                                                                & Baseline-2             \\ \cline{3-11} 
                                                                        &                                                                      & 16         & 12         & 4          & 0           & LRW              & No                  & 200                                                                          & 8.60                                                                & Baseline-2             \\ \cline{3-11} 
                                                                        &                                                                      & 16         & 12         & 4          & 0           & LRW              & Yes                 & 1000                                                                         & 16.63                                                               & Baseline-2             \\ \hline
\end{tabular}
\end{table*}

\subsection{Design Space}
In this section, we evaluate the proposed architecture for various parameter combinations. The parameters in our work are the cache size, main memory size, and sizes of K, M, N, and WC. We intend to allow the end user to choose a suitable architecture with an appropriate parameter size.

As shown in the tables \ref{tab1} and \ref{tab2}, we used the same experimental setup and latency/energy values for NVM. Our design depends on the number of power failures that occur. We performed experiments to observe the energy gains when the number of power failures increased from 500 to 1000 and decreased from 500 to 200. Usually, we introduce a power failure for every 2 million instructions, which means that in executing 1 billion instructions, we experience 500 power failures. We introduce a power failure for every 1 million instructions to increase the number of failures from 500 to 1000. Similarly, if we introduce a power failure every 5 million instructions, the total number of power failures becomes 200.

So, our design space depends on (cache\_size, Memory\_size, K, M, N, WC, Replacement\_policies, BR, number of power failures). If we fix cache, memory, and WC sizes w.r.t our experimental setup, our design space becomes large enough. Let's take an instance to see how large our design space will become. For K=16, possible K values are 16. If K=16, possible (M, N) pairs are 16*16. For BR, possible conditions are 2 (BR-enabled or not). For replacement policies, the possible ways are 2 (LRW or LFW). The number of power failures; the possible cases are 4 (as per the experiments we performed) if we combine all these parameters for calculating the design space size (16*16*16*2*2*4) equals 65,536 design choices for given cache size, main-memory size, WC, and K=16. 

We can notice from table \ref{tabc2} that only one design choice performs better in all of the given combinations. For the given cache size, main-memory size, and K=16, we find that the combination of M=12, N=4, WC=6, BR-enabled, and the number of power failures=1000 is preferable to all other 65,536 design choices.

\section{Conclusions} \label{p6}
In this paper, we proposed an NVM-based architecture. Using the proposed DBT and WBQ, we see fewer writes to STT-RAM (LLC) and PCM (main memory). The proposed architecture decreases STT-RAM writes by 18.97\% and PCM writes by 10.66\% compared with baseline-1 architecture. As a result, we have decreased energy consumption by about 17.56\%. However, the proposed architecture has 5.10\% execution overhead and 4.93\% energy overhead compared to baseline-2 architecture under the stable power supply. We also compared the existing checkpointing policies with the proposed architecture. We introduced an STT-RAM-based backup region at LLC that helps for backup from L1 during a power failure. We also evaluated and analyzed the unified NVM architecture with the proposed architecture. We explored various design spaces to determine how our proposed architecture behaves when changing parameter sizes.

\section{Acknowledgement}
This work is supported by the grant received from DST, Govt. of India for the Technology Innovation Hub at the IIT Ropar in the framework of the National Mission on Interdisciplinary Cyber-Physical Systems.

\bibliographystyle{unsrt}
\bibliography{main}{}

\begin{thebibliography}{10}

\bibitem{big}
H{\^e}ri{\c{s}} Golp{\^\i}ra, Syed Abdul~Rehman Khan, and Sina Safaeipour.
\newblock A review of logistics internet-of-things: Current trends and scope
  for future research.
\newblock {\em Journal of Industrial Information Integration}, page 100194,
  2021.

\bibitem{lifetime}
Xiaosong Hu, Le~Xu, Xianke Lin, and Michael Pecht.
\newblock Battery lifetime prognostics.
\newblock {\em Joule}, 4(2):310--346, 2020.

\bibitem{energy12}
Dong Ma, Guohao Lan, Mahbub Hassan, Wen Hu, and Sajal~K Das.
\newblock Sensing, computing, and communications for energy harvesting iots: A
  survey.
\newblock {\em IEEE Communications Surveys \& Tutorials}, 22(2):1222--1250,
  2019.

\bibitem{int3}
Josiah Hester and Jacob Sorber.
\newblock The future of sensing is batteryless, intermittent, and awesome.
\newblock In {\em Proceedings of the 15th ACM conference on embedded network
  sensor systems}, pages 1--6, 2017.

\bibitem{intermittent}
Brandon Lucia, Vignesh Balaji, Alexei Colin, Kiwan Maeng, and Emily Ruppel.
\newblock Intermittent computing: Challenges and opportunities.
\newblock {\em 2nd Summit on Advances in Programming Languages (SNAPL 2017)},
  2017.

\bibitem{forget}
Fang Su, Yongpan Liu, Yiqun Wang, and Huazhong Yang.
\newblock A ferroelectric nonvolatile processor with 46 $ \backslash \mu $ s
  system-level wake-up time and 14$ \backslash \mu $ s sleep time for energy
  harvesting applications.
\newblock {\em IEEE Transactions on Circuits and Systems I: Regular Papers},
  64(3):596--607, 2016.

\bibitem{r8}
Albert Lee, Chieh-Pu Lo, et~al.
\newblock A reram-based nonvolatile flip-flop with self-write-termination
  scheme for frequent-off fast-wake-up nonvolatile processors.
\newblock {\em IEEE Journal of Solid-State Circuits}, 52(8):2194--2207, 2017.

\bibitem{r41}
Hrishikesh Jayakumar, Arnab Raha, and Vijay Raghunathan.
\newblock Energy-aware memory mapping for hybrid fram-sram mcus in iot edge
  devices.
\newblock In {\em 2016 29th International Conference on VLSI Design and 2016
  15th International Conference on Embedded Systems (VLSID)}, pages 264--269.
  IEEE, 2016.

\bibitem{capac}
Attapong Mamen and Uthane Supatti.
\newblock A survey of hybrid energy storage systems applied for intermittent
  renewable energy systems.
\newblock In {\em 2017 14th ECTI-CON}, pages 729--732. IEEE, 2017.

\bibitem{int2}
Milijana Surbatovich, Brandon Lucia, and Limin Jia.
\newblock Towards a formal foundation of intermittent computing.
\newblock {\em Proceedings of the ACM on Programming Languages},
  4(OOPSLA):1--31, 2020.

\bibitem{stt1}
Sheel~Sindhu Manohar and Hemangee~K Kapoor.
\newblock Capmig: Coherence aware block placement and migration in
  multi-retention stt-ram caches.
\newblock {\em IEEE TCAD}, 2022.

\bibitem{pcm}
Sparsh Mittal and Jeffrey~S Vetter.
\newblock A survey of software techniques for using non-volatile memories for
  storage and main memory systems.
\newblock {\em IEEE Transactions on Parallel and Distributed Systems},
  27(5):1537--1550, 2015.

\bibitem{msp}
Texas Instruments.
\newblock Msp430fr5969 launchpad development kit, 2018.

\bibitem{rev2}
Sandeep Thirumala, Arnab Raha, Sumeet Gupta, and Vijay Raghunathan.
\newblock Exploring the design of energy-efficient intermittently powered
  systems using reconfigurable ferroelectric transistors.
\newblock {\em IEEE Transactions on Very Large Scale Integration (VLSI)
  Systems}, 30(4):365--378, 2021.

\bibitem{reg2}
Aika Kamei, Hideharu Amano, Takuya Kojima, Daiki Yokoyama, Kimiyoshi Usami,
  Keizo Hiraga, Kenta Suzuki, and Kazuhiro Bessho.
\newblock A variation-aware mtj store energy estimation model for edge devices
  with verify-and-retryable nonvolatile flip-flops.
\newblock {\em IEEE Transactions on Very Large Scale Integration (VLSI)
  Systems}, 2023.

\bibitem{xie}
Mimi Xie, Chen Pan, and Chun~Jason Xue.
\newblock A novel stt-ram-based hybrid cache for intermittently powered
  processors in iot devices.
\newblock {\em IEEE Micro}, 39(1):24--32, 2018.

\bibitem{llc1}
Kunal Korgaonkar, Ishwar Bhati, Huichu Liu, Jayesh Gaur, Sasikanth Manipatruni,
  Sreenivas Subramoney, Tanay Karnik, Steven Swanson, Ian Young, and Hong Wang.
\newblock Density tradeoffs of non-volatile memory as a replacement for sram
  based last level cache.
\newblock In {\em 2018 ACM/IEEE 45th Annual International Symposium on Computer
  Architecture (ISCA)}, pages 315--327. IEEE, 2018.

\bibitem{llc2}
Candace Walden, Devesh Singh, Meenatchi Jagasivamani, Shang Li, Luyi Kang,
  Mehdi Asnaashari, Sylvain Dubois, Bruce Jacob, and Donald Yeung.
\newblock Monolithically integrating non-volatile main memory over the
  last-level cache.
\newblock {\em ACM Transactions on Architecture and Code Optimization (TACO)},
  18(4):1--26, 2021.

\bibitem{choi}
Ju-Hee Choi and Gi-Ho Park.
\newblock Nvm way allocation scheme to reduce nvm writes for hybrid cache
  architecture in chip-multiprocessors.
\newblock {\em IEEE Transactions on Parallel and Distributed Systems},
  28(10):2896--2910, 2017.

\bibitem{r2}
Benjamin~C Lee, Engin Ipek, Onur Mutlu, and Doug Burger.
\newblock Architecting phase change memory as a scalable dram alternative.
\newblock In {\em Proceedings of the 36th annual international symposium on
  Computer architecture}, pages 2--13, 2009.

\bibitem{r3}
Moinuddin~K Qureshi, Michele~M Franceschini, and Luis~A Lastras-Montano.
\newblock Improving read performance of phase change memories via write
  cancellation and write pausing.
\newblock In {\em HPCA-16 2010 The Sixteenth International Symposium on
  High-Performance Computer Architecture}, pages 1--11. IEEE, 2010.

\bibitem{backup}
Hrishikesh Jayakumar, Arnab Raha, Jacob~R Stevens, and Vijay Raghunathan.
\newblock Energy-aware memory mapping for hybrid fram-sram mcus in
  intermittently-powered iot devices.
\newblock {\em ACM Transactions on Embedded Computing Systems (TECS)},
  16(3):1--23, 2017.

\bibitem{check2}
Saad Ahmed, Naveed~Anwar Bhatti, Muhammad~Hamad Alizai, Junaid~Haroon Siddiqui,
  and Luca Mottola.
\newblock Efficient intermittent computing with differential checkpointing.
\newblock In {\em Proceedings of the 20th ACM SIGPLAN/SIGBED International
  Conference on Languages, Compilers, and Tools for Embedded Systems}, pages
  70--81, 2019.

\bibitem{check3}
Saad Ahmed, Naveed~Anwar Bhatti, Muhammad~Hamad Alizai, Junaid~Haroon Siddiqui,
  and Luca Mottola.
\newblock Fast and energy-efficient state checkpointing for intermittent
  computing.
\newblock {\em ACM Transactions on Embedded Computing Systems (TECS)},
  19(6):1--27, 2020.

\bibitem{rev21}
Sandeep~Krishna Thirumala, Arnab Raha, Vijay Raghunathan, and Sumeet~Kumar
  Gupta.
\newblock Ips-cim: Enhancing energy efficiency of intermittently-powered
  systems with compute-in-memory.
\newblock In {\em 2020 IEEE 38th International Conference on Computer Design
  (ICCD)}, pages 368--376. IEEE, 2020.

\bibitem{r9}
Fan Zhang, Yanqing Zhang, et~al.
\newblock A batteryless 19$\mu$w mics/ism-band energy harvesting body area
  sensor node soc.
\newblock In {\em 2012 IEEE International Solid-State Circuits Conference},
  pages 298--300. IEEE, 2012.

\bibitem{r10}
Domenico Balsamo, Anup Das, et~al.
\newblock Graceful performance modulation for power-neutral transient computing
  systems.
\newblock {\em IEEE TCAD}, 35(5):738--749, 2016.

\bibitem{r11}
Hehe Li, Yongpan Liu, Qinghang Zhao, et~al.
\newblock An energy efficient backup scheme with low inrush current for
  nonvolatile sram in energy harvesting sensor nodes.
\newblock In {\em 2015 DATE}, pages 7--12. IEEE, 2015.

\bibitem{r12}
Xiao Sheng, Yiqun Wang, Yongpan Liu, and Huazhong Yang.
\newblock Spac: A segment-based parallel compression for backup acceleration in
  nonvolatile processors.
\newblock In {\em 2013 DATE}, pages 865--868. IEEE, 2013.

\bibitem{r13}
Yiqun Wang, Yongpan Liu, Shuangchen Li, et~al.
\newblock Pacc: A parallel compare and compress codec for area reduction in
  nonvolatile processors.
\newblock {\em IEEE TVLSI}, 22(7):1491--1505, 2013.

\bibitem{cache2n}
Yu-Ting Chen, Jason Cong, Hui Huang, Bin Liu, Chunyue Liu, Miodrag Potkonjak,
  and Glenn Reinman.
\newblock Dynamically reconfigurable hybrid cache: An energy-efficient
  last-level cache design.
\newblock In {\em 2012 Design, Automation \& Test in Europe Conference \&
  Exhibition (DATE)}, pages 45--50. IEEE, 2012.

\bibitem{n3}
Feng Chen, Song Jiang, and Xiaodong Zhang.
\newblock Smartsaver: Turning flash drive into a disk energy saver for mobile
  computers.
\newblock In {\em Proceedings of the 2006 international symposium on Low power
  electronics and design}, pages 412--417, 2006.

\bibitem{n11}
Taeho Kgil, David Roberts, and Trevor Mudge.
\newblock Improving nand flash based disk caches.
\newblock In {\em 2008 International Symposium on Computer Architecture}, pages
  327--338. IEEE, 2008.

\bibitem{n26}
Michael Wu and Willy Zwaenepoel.
\newblock envy: a non-volatile, main memory storage system.
\newblock {\em ACM SIGOPS Operating Systems Review}, 28(5):86--97, 1994.

\bibitem{n108}
Jalil Boukhobza, St{\'e}phane Rubini, Renhai Chen, and Zili Shao.
\newblock Emerging nvm: A survey on architectural integration and research
  challenges.
\newblock {\em ACM Transactions on Design Automation of Electronic Systems
  (TODAES)}, 23(2):1--32, 2017.

\bibitem{cache1n}
Yiran Chen, Weng-Fai Wong, Hai Li, and Cheng-Kok Koh.
\newblock Processor caches built using multi-level spin-transfer torque ram
  cells.
\newblock In {\em IEEE/ACM International Symposium on Low Power Electronics and
  Design}, pages 73--78. IEEE, 2011.

\bibitem{r1}
Zhe Wang, Shuchang Shan, Ting Cao, et~al.
\newblock Wade: Writeback-aware dynamic cache management for nvm-based main
  memory system.
\newblock {\em ACM TACO}, 10(4):1--21, 2013.

\bibitem{13}
Nathan Binkert, Bradford Beckmann, et~al.
\newblock The gem5 simulator.
\newblock {\em ACM SIGARCH computer architecture news}, 39(2):1--7, 2011.

\bibitem{14}
Matthew~R Guthaus et~al.
\newblock Mibench: A free, commercially representative embedded benchmark
  suite.
\newblock In {\em Proceedings of the fourth annual IEEE international workshop
  on workload characterization}, pages 3--14. IEEE, IEEE, 2001.

\bibitem{25}
Xiangyu Dong, Cong Xu, Yuan Xie, and Norman~P Jouppi.
\newblock Nvsim: A circuit-level performance, energy, and area model for
  emerging nonvolatile memory.
\newblock {\em IEEE TCAD}, 31(7):994--1007, 2012.

\end{thebibliography}

\vfill

\end{document}